\documentclass{article}

\usepackage{microtype}
\usepackage{graphicx}
\usepackage{amsmath}
\DeclareMathOperator{\Com}{Com}
\usepackage{subfigure}
\usepackage{booktabs} % for professional tables

\usepackage{hyperref}

\usepackage[accepted]{mlsys2025}

\usepackage{hyperref}
\usepackage{url}
\usepackage{natbib}
\usepackage{amsmath,amssymb,mathtools, amsthm, bm, amsfonts}
\usepackage{algorithmic}
\usepackage{algorithm}
\usepackage{graphicx}
\usepackage{textcomp}
\usepackage{booktabs}
\usepackage{multirow}
\usepackage{siunitx} 
\usepackage{caption}
\usepackage{makecell}

\usepackage{dsfont}    
\usepackage{microtype} 
\usepackage{amsfonts} % For \mathbb{R} and other math symbols
\usepackage{xspace}
\usepackage{enumitem}
\usepackage{array}
\usepackage{tabularx}
\usepackage{graphicx}  % make sure this is in your preamble
\mlsystitlerunning{ZK-APEX\xspace: Zero-Knowledge Approximate Personalized Unlearning with EXecutable Proofs}

\begin{document}

\newcommand{\name}{\textsf{ZK\text{-}APEX}\xspace}
\newcommand{\forgetset}{\textit{forget-set}\xspace}
\newcommand{\forgetclass}{\textit{forget class}\xspace}
\newcommand{\retainset}{\textit{retain-set}\xspace}
\newcommand{\colonpoint}[1]{\noindent\textbf{#1:}~}
\newcommand{\point}[1]{\par\smallskip{\noindent\textbf{#1.}~}}

\twocolumn[
\mlsystitle{\name: Zero-Knowledge Approximate Personalized Unlearning with EXecutable Proofs}

% unless you've provided the [accepted] option to the mlsys2025


% \mlsyssetsymbol{equal}{*}

\begin{mlsysauthorlist}
\mlsysauthor{Mohammad M Maheri}{to}
\mlsysauthor{Sunil Cotterill}{to}
\mlsysauthor{Alex Davidson}{goo}
\mlsysauthor{Hamed Haddadi}{to}
\end{mlsysauthorlist}

\mlsysaffiliation{to}{Imperial College London}
\mlsysaffiliation{goo}{LASIGE, Faculdade de Ciências, Universidade de Lisboa}
% \mlsysaffiliation{ed}{School of Computation, University of Edenborrow, Edenborrow, United Kingdom}

\mlsyscorrespondingauthor{Mohammad M Maheri}{m.maheri23@imperial.ac.uk}
% \mlsyscorrespondingauthor{Eee Pppp}{ep@eden.co.uk}

\mlsyskeywords{Machine Learning, Deep Learning, Machine Unlearning, Personalization, Zero-knowledge Proofs}

\vskip 0.3in

\begin{abstract}

Machine unlearning removes the influence of specified data from trained models to satisfy privacy, copyright, and safety requirements (e.g., the “right to be forgotten”). In practice, providers distribute a global model to edge devices, that each locally personalize the model based on their private data. 
However, since clients may ignore or falsify deletion requests, providers must verify correct unlearning for these distributed models, without accessing private parameters. This is particularly challenging for personalized models, which must forget designated samples without degrading local utility, while ensuring that verification remains efficient and scalable on resource-constrained edge devices.

We formalize personalized unlearning and develop a zero-shot approximate unlearning algorithm that works directly on the personalized model without retraining. Our novel method, \name, combines provider-side sparse masking for targeted removal with client-side Group-OBS compensation computed from a block-wise empirical Fisher. This technique yields a curvature-aware update designed for low-overhead execution and proof generation. Using modern Halo2 ZK-SNARKs, we prove operator compliance by showing that the unlearned model exactly matches the committed output of the prescribed transformation, without revealing personalized model parameters or data.

% On ViT models for ImageNet-1k and ImageNet-Sketch, our approach achieves effective forgetting while recovering 81\% of personalization accuracy. Proofs complete in $\approx$2 hours, which is over $10^4\times$ faster than training-based verification, with peak memory under 1 GB, and proof sizes below 24 MB. Our work establishes the first verifiable personalized unlearning framework feasible for deployment on edge devices.

% On Vision Transformer (ViT) classification models, our approach recovers approximately 99\% Top-1 personalization accuracy while enforcing effective forgetting. We further evaluate the unlearning algorithm on a generative model, OPT125M, trained on the CodeParrot code dataset, achieving $\sim$70\% recovery of original accuracy. ZK-SNARK proof generation for the ViT case completes in $\approx$2~hours, which is more than $10^7\times$ faster than retraining based verification, with peak memory under 0.7~GB and proof sizes about 400~MB. Together, these results establish the first verifiable personalized unlearning framework practical for deployment on resource constrained edge devices.

On Vision Transformer (ViT) classification models, our approach recovers approximately 99\% Top-1 personalization accuracy while enforcing effective forgetting. We further evaluate the unlearning algorithm on a generative model, the autoregressive language model OPT125M, trained on the CodeParrot code dataset, achieving $\sim$70\% recovery of original accuracy. ZK-SNARK proof generation for the ViT case completes in $\approx$2~hours, which is more than $10^7\times$ faster than retraining based verification, with peak memory under 0.7~GB and proof sizes about 400~MB. Together, these results establish the first verifiable personalized unlearning framework practical for deployment on resource constrained edge devices.

\end{abstract}
]



\printAffiliationsAndNotice{}  % leave blank if no need to mention equal contribution
% \printAffiliationsAndNotice{\mlsysEqualContribution} % otherwise use the standard text.

\section{Introduction}

Machine unlearning aims to remove the influence of specific data points from trained models, thereby aligning machine learning systems with privacy regulations such as General Data Protection Regulation (GDPR)~\cite{mantelero2013eu} and 
California Privacy Rights Act (CPRA)~\cite{harding2019understanding}, which grant individuals the right to be forgotten~\cite{regulation2016regulation,cao2015towards}. Beyond privacy compliance, unlearning is also required for mitigating biases~\cite{liu2025disentangling}, correcting corrupted data, or retracting samples obtained under invalid consent~\cite{li2025machine}.
% In many practical deployments, the entity requesting unlearning and the party executing it operate under a trustless setting. For example, a model provider may distribute a pretrained model to multiple clients, each of whom personalizes it using locally-collected private data. When the provider later requests the removal of certain samples, they must be able to verify that each client has correctly performed the unlearning, while clients must avoid disclosing their personalized models, as these encode sensitive information over said private data~\cite{balle2022reconstructing,leino2020stolen}.
In many practical deployments, such as smartphone keyboards~\cite{hard2018federated,singhal2021federated}, photo categorization models (for example Google Photos and Apple Photos)~\cite{gunter2024apple}, and voice assistants~\cite{du2024communication}, the model provider, typically a company, distributes a pretrained model to edge client devices. Each device personalizes the model on locally collected private data, as recently demonstrated in Apple Intelligence~\cite{gunter2024apple}, which enables large-scale on-device adaptation through lightweight foundation models. When the provider later requests the removal of certain samples, it must verify that each client has executed the unlearning correctly, while the clients must not expose their personalized models, which contain sensitive private information~\cite{balle2022reconstructing,leino2020stolen}.
This creates a verification challenge: the client, who holds a locally personalized model, must convince the model provider that the specified data has been erased, without exposing the underlying model parameters or private samples. The problem is further complicated in edge environments, where storage, compute, and communication constraints make full retraining or model sharing infeasible. Consequently, there is a growing need for privacy-preserving verification mechanisms that can attest to correct unlearning on-device.
% Recently, zero-knowledge succinct non-interactive arguments of knowledge (ZK-SNARKs) have emerged as a powerful cryptographic primitive for verifiable computation, including neural network inference~\cite{maheri2025telesparse,t}. In essence, ZK-SNARKs allow a prover (e.g., an edge device) to generate a compact proof that a computation—such as an unlearning update—was executed correctly, which can then be publicly verified without revealing any intermediate values and model parameters.
A natural direction for enabling such verifiable unlearning is the use of zero-knowledge succinct non-interactive arguments of knowledge (ZK-SNARKs), a cryptographic primitive that enables verifiable computation without revealing private information. ZK-SNARKs allow a prover (e.g., an edge device) to generate a compact proof that a computation, such as an unlearning update, has been performed correctly, while the verifier (e.g., the model provider) can efficiently check its validity without accessing the personalized model parameters or intermediate values~\cite{kang2022scaling,maheri2025telesparse,rabanser2025confidential}.
 
 % \textcolor{red}{[Describe Personalized scenario and it challenges $\to$ What is the goal]} \\
 % However, personalized model unlearning on edge devices introduces unique challenges. Unlike centralized settings where the model provider can compute and distribute unlearning updates, applying such updates directly to personalized models can severely degrade their tailored performance, undermining the benefits of personalization. Furthermore, naively sending the raw unlearning data (i.e., the \forgetset) to the client and requesting proof of unlearning is infeasible: 
 % i) it requires clients to generate ZK-SNARK proofs for full unlearning algorithms—such as multi-epoch gradient ascent or retraining on the retained dataset—which demands prohibitively high computation and memory, making it impractical for edge devices. Generating ZK-SNARK proofs for such heavy procedures is particularly unrealistic, as proof generation for even moderate-scale training remains orders of magnitude more expensive than inference-level proofs~\cite{sun2024zkdl}, and 
 % ii) it exposes the raw \forgetset , compromising its privacy. 
 % These limitations highlight the need for a new approach that enables efficient and privacy-preserving verification of unlearning on locally personalized models, without revealing sensitive data or compromising model utility. In this work, we address this gap by designing a ZK-SNARK-friendly approximate unlearning procedure tailored for personalized models on edge devices.
However, performing unlearning in personalized settings on edge devices introduces distinctive challenges. Unlike centralized unlearning frameworks~\cite{kurmanji2023towards,thudi2022unrolling,chundawat2023can,chundawat2023zero,jia2023model}, where the model provider can compute and distribute global unlearning updates, directly applying such updates to locally personalized models often degrades their task-specific adaptation. Moreover, naive strategies that require the provider to send the \forgetset to each client and demand proof of its removal are infeasible for several reasons.
% First, generating zero-knowledge proofs for full unlearning procedures—such as multi-epoch gradient ascent, retraining, or optimization over the retained dataset—incurs prohibitive computational and memory overhead. Proof generation for these complex updates remains several orders of magnitude more expensive than inference-level verification~\cite{sun2024zkdl}, rendering them impractical and especially unsuitable for edge hardware.
First, generating zero-knowledge proofs (ZKP) for complete unlearning procedures such as multi-epoch gradient ascent~\cite{graves2021amnesiac,thudi2022unrolling}, retraining~\cite{bourtoule2021machine,eisenhofer2025verifiable,yu2023split,xia2025edge}, or fine-tuning over the retained dataset~\cite{hong2024dissecting} imposes significant computational and memory overhead~\cite{qu2025zkgpt,maheri2025telesparse,sun2024zkdl}. Compared to inference, producing ZKPs for training is substantially more complex~\cite{qu2025zkgpt,sun2024zkdl}. 
Prior studies~\cite{abbaszadeh2024zero,garg2023experimenting,waiwitlikhit2024trustless} have explored proofs of training for different machine learning models. However, these methods remain prohibitively expensive and impractical for real-world deployment, especially on edge devices.
Second, sharing the raw \forgetset with clients is undesirable, as these samples are typically sensitive in unlearning applications~\cite{xia2025edge,nguyen2025survey}.
These limitations motivate a new paradigm for verifiable personalized unlearning that is both computationally tractable and privacy-preserving. 
% The key goal is to enable clients to produce succinct proofs attesting that unlearning has been faithfully performed on their personalized models, without revealing private data or degrading model utility. In this work, we address this challenge by proposing a ZK-SNARK-compatible approximate unlearning framework specifically designed for personalized edge models.
The key objective is to develop an unlearning mechanism that (i) effectively removes the influence of designated samples from personalized models without sacrificing their local performance, and (ii) remains compatible with ZKP, ensuring that the verification of unlearning is computationally tractable for edge devices. In this work, we address this challenge by introducing a ZK-friendly approximate unlearning framework tailored for personalized models in resource-constrained edge environments.

% \textcolor{red}{Contributions of the paper in order to address the mentioned challenges}\\

Recent studies have shown that information in deep networks tends to localize within specific neurons or filters~\cite{ghorbani2020neuron,lin2020hrank}. This observation suggests that removing or masking a small subset of highly influential weights can effectively erase the knowledge associated with targeted samples. Building on this insight, we design our unlearning request as a saliency-based pruning operation, where each weight is scored according to its contribution to the \forgetset, using first- and second-order statistics such as gradients and curvature. Masking the weights with the highest scores removes the influence of the forget samples from the model~\cite{jia2023model,hong2024dissecting}.  
However, directly applying such a mask to a personalized model substantially degrades its accuracy, since personalization relies on locally adapted features that often overlap with those associated with the \forgetset. To mitigate this, we formulate a compensation step grounded in the Optimal Brain Surgeon (OBS) framework~\cite{lecun1989optimal,kurtic2022optimal}, which computes a second-order weight adjustment that restores performance on the personalized data while maintaining high loss on the forget domain. This two-part procedure achieves the first goal of effective unlearning on personalized models while preserving their utility.
The second goal is verifiability under zero knowledge (ZK). Our algorithm is inherently ZK-friendly because it is \emph{zero-shot}, performing no stochastic training or iterative optimization after applying the unlearning transformation, and thus avoids the randomness of SGD, which has been shown to enable forging attacks in verifiable unlearning~\cite{DBLP:conf/icml/ZhangCSL24}. Instead, the client only needs to prove that (i) the specified mask correctly zeroed the targeted parameters and (ii) the compensation weights were computed according to the prescribed rule from committed inputs, including the Fisher information matrix of the personalized model (evaluated on the personalization set) and the public mask. These operations reduce to sparse matrix–vector computations that can be efficiently verified within a ZK-SNARK circuit.
% As a result, the proposed method enables practical and privacy-preserving verification of personalized unlearning across diverse deployment scenarios, with proof generation that is even more computationally efficient than typical ZK-SNARK-based inference verification.
As a result, we introduce \name, 
\emph{Zero-Knowledge Approximate Personalized Unlearning with EXecutable Proofs}, 
which enables practical and privacy-preserving verification of personalized unlearning 
across diverse deployment scenarios.  
\name achieves proof generation that is even more computationally efficient 
than typical ZK-SNARK-based inference verification on a single sample.

\noindent \textbf{Contributions.}
% In this paper, we make the following contributions.
We present the following key contributions.
\begin{itemize}[leftmargin=*, nosep]
\item \textbf{Novel formulation of personalized unlearning.}
We introduce and formalize the personalized unlearning problem, which removes the influence of \forgetset samples from locally adapted models while preserving user-specific performance. This formulation goes beyond conventional unlearning by modeling realistic scenarios where models personalized on private data must remain verifiable under trustless and privacy-sensitive conditions.

% \item \textbf{Principled curvature-based and ZK-friendly unlearning method.} 
% We propose a novel framework that eliminates \forgetset information through curvature-aware masking and Optimal Brain Surgeon compensation, \emph{leveraged here for the first time in the context of machine unlearning}.
% The method is grounded in a theoretically motivated formulation that balances forgetting efficacy and personalization retention, leveraging second-order structure to limit accuracy degradation.
% It remains computationally tractable and naturally compatible with zero-knowledge proof generation.

% We propose a novel framework that removes \forgetset information through curvature-aware masking and OBS compensation, \emph{leveraged for the first time in the context of machine unlearning}.
% Grounded in a theoretically motivated formulation, the method balances forgetting efficacy and personalization retention by exploiting second-order curvature structure while ensuring tractability.
% Its closed-form, linear operation design makes it computationally efficient and naturally compatible with zero-knowledge proof generation.
\item \textbf{Principled curvature-based and ZK-friendly unlearning method.}
We propose a framework that removes \forgetset information through curvature-aware masking and OBS compensation, \emph{leveraged for the first time in machine unlearning}.
Grounded in a principled formulation, the method balances forgetting efficacy and personalization retention by exploiting second-order curvature while ensuring tractability.
Its closed-form linear design makes it computationally efficient and naturally compatible with ZKP generation.

\item \textbf{Verifiable unlearning via efficient zero knowledge.}  
We design a ZK-SNARK-based verification system that certifies correct execution of the unlearning procedure without exposing private data or model parameters. The proof construction only involves sparse and linear operations, leading to proving costs substantially lower than prior ZK-SNARK approaches for training and even more efficient than inference-level verification.  

% \item \textbf{Comprehensive empirical validation.}  
% We evaluate our method across diverse personalized learning scenarios, including classification tasks with Vision Transformers (ViT) and generative tasks with large language models (LLM). We assess unlearning effectiveness, retention of personalized accuracy, and ZK-SNARK proof-generation efficiency using the Halo2 framework. Additional experiments on mobile devices measure on-device proving latency and computational overhead, confirming that the proposed approach achieves practical and resource-efficient verifiable unlearning.
\item \textbf{Comprehensive empirical validation.}
We evaluate our method across personalized learning scenarios, including classification with Vision Transformers (ViT) and generation with large language models (LLM). We measure unlearning efficacy, personalization retention, and ZK-SNARK proof efficiency under Halo2, and further test on mobile devices to confirm practical, resource-efficient verifiable unlearning.
\end{itemize}

Our open-source implementation is available at \url{https://github.com/mammadmaheri7/ZK-APEX}.

% \textcolor{red}{TODO: explain why non-interactive ZKP and specifically ZK-SNARK is a good design choice for the problem}

% \textcolor{red}{why normal unlearning does not works in this problem:}
% Compared to inference, generating zero-knowledge proofs for training is considerably more challenging~\cite{qu2025zkgpt}. Prior studies~\cite{abbaszadeh2024zero,garg2023experimenting,waiwitlikhit2024trustless,shamsabadi2022confidential} have investigated proofs of training for different machine learning models, but existing techniques remain prohibitively expensive, introducing substantial computational and memory overheads.

% \textcolor{red}{TODO: mention this work is orthogonal to speedup (parallelization) of proving of zero knowledge proof schema as it would take benefit of them but for fair comparison among different unlearning algorithms all of them implemented under same model ZKP of halo2} \\

% \textcolor{red}{TODO: explain distinction with "Verifiable and Provably Secure Machine Unlearning"}

\section{Related Works and Background}

\paragraph{Approximate Machine Unlearning.}

Machine unlearning was first introduced to ensure that trained models could forget specific data as if it had never been used~\cite{cao2015towards,bourtoule2021machine,graves2021amnesiac}.
Approaches such as SISA~\cite{bourtoule2021machine} and CAUSE~\cite{xia2025edge} 
% and Amnesiac Unlearning~\cite{graves2021amnesiac} 
achieve this through retraining on data partitions or maintaining multiple checkpoints of the training process.
While these methods guarantee deletion in theory, they depend on access to the full dataset and impose substantial storage and computation costs, making them impractical for large-scale or distributed deployments.

To address these limitations, approximate unlearning methods modify the trained model directly to emulate the effect of retraining~\cite{kurmanji2023towards,golatkar2020forgetting,chundawat2023can,thudi2022unrolling}.
A common formulation is class unlearning, where the goal is to erase representations associated with particular semantic categories, while preserving the model’s ability to generalize to the remaining classes~\cite{chundawat2023zero,seo2025revisiting,fan2024challenging}.
The main challenge lies in suppressing class-specific information without damaging the broader structure of shared features, as excessive removal can harm retained-class accuracy.

More recently, studies have shown that neural representations are spatially and structurally localized~\cite{frantar2023sparsegpt,meng2022locating}, motivating sparsity-based unlearning approaches that identify and mask a small set of influential weights to remove targeted information~\cite{fan2023salun,jia2023model,pochinkov2024dissecting}.
Such methods have been successfully applied to both discriminative architectures~\cite{jia2023model} and generative models~\cite{fan2023salun,pochinkov2024dissecting} such as large language models, demonstrating that selective parameter masking can effectively achieve forgetting, while preserving global performance.
Despite their efficiency, existing approximate unlearning approaches generally assume a centralized setup, where the model provider performs unlearning on a shared global model using the \forgetset data.
In contrast, our work considers unlearning in personalized models that have been locally adapted using private client data, aiming to remove provider-specified class information while retaining personalization-specific knowledge.

% \subsection{Verifiable Machine Learning/Unlearning}
% \textcolor{red}{Stress on the importance of non-interactive protocols in most ML application}

% \textcolor{red}{two categories: backdoor verification, reproducing verification $\to$ both of them not applicable to personalized scenario}

\paragraph{Verifiable Machine Learning and Unlearning.}
% ZKPs~\cite{goldreich1994definitions} enable verifiable machine learning by allowing a prover to convince a verifier that a computation was correctly executed without revealing private inputs or intermediate states (the witnesses).
% Among modern proving systems, succinct non-interactive arguments of knowledge (SNARKs)~\cite{ITCS:BCCT12,kilian1992note,micali2000computationally} are widely adopted for ML verification due to their short proofs, fast verification, non-interactivity, and modular circuit design that can be partitioned across model layers~\cite{maheri2025telesparse,liu2021zkcnn,lee2024vcnn,sun2024zkllm,zcash-halo2}.
% These properties make ZK-SNARKs suitable for large-scale ML auditing.
% Nonetheless, proof generation remains computationally intensive, with significant time and memory overheads for large models~\cite{kang2022scaling}.

ZKPs~\cite{goldreich1994definitions} enable verifiable ML by allowing a prover to demonstrate correct computation without revealing private inputs or intermediate states.
ZK-SNARKs~\cite{ITCS:BCCT12,kilian1992note,micali2000computationally} are widely used for ML verification owing to their short proofs, fast verification, and modular circuits that can be partitioned across layers~\cite{maheri2025telesparse,liu2021zkcnn,lee2024vcnn,sun2024zkllm,zcash-halo2}.
While suitable for large-scale auditing, proof generation remains costly in time and memory~\cite{kang2022scaling}.

Verification of machine unlearning ensures that models have genuinely forgotten target data.
Existing methods include backdoor and sensitivity-based verification, which use poisoned or sensitivity of samples to test deletion~\cite{sommer2022athena,gao2024verifi,guo2023verifying,zhou2025truvrf}, but can be bypassed by a dishonest prover, and reproducing verification, which replays unlearning traces via proofs of learning~\cite{thudi2022necessity,weng2024proof,eisenhofer2025verifiable} but remains computationally heavy.
% Recent cryptographic frameworks instantiate proofs of unlearning with SNARKs and hash chains~\cite{eisenhofer2025verifiable}, yet these verify exact unlearning (full retraining), making proof generation impractical.
% Moreover, such methods are not fully robust to the stochasticity inherent in retraining, as the randomness of SGD can be exploited to forge seemingly valid proofs~\cite{DBLP:conf/icml/ZhangCSL24}.
Recent cryptographic frameworks instantiate proofs of unlearning with SNARKs and hash chains~\cite{eisenhofer2025verifiable}, yet these verify exact unlearning (full retraining), making proof generation impractical.
Furthermore, they are vulnerable to the stochasticity of SGD~\cite{DBLP:conf/icml/ZhangCSL24}: an adversary can exploit minibatch randomness to retrain on retained samples whose gradients mimic those of the removed data, producing an apparently valid proof even though the model still encodes the \forgetset information.
This motivates ZK-friendly approximate unlearning methods that achieve verifiability without retraining.

\subsection{Optimal Brain Surgeon}
% \textcolor{red}{Introduce the obs and related works as compensation of model pruning with goal of fast leaning $\to$ Need approximation of hessian $\to$ fisher information}

\paragraph{Optimal Brain Surgeon and Fisher-based compensation.}
Second-order pruning methods such as Optimal Brain Damage~\cite{lecun1989optimal} and Optimal Brain Surgeon~\cite{hassibi1993optimal} introduced a principled approach to compensate for parameter removal by modeling loss curvature.
By expanding the loss near an optimum, these methods derive closed-form updates that minimally affect performance after zeroing selected weights.
Later work proposed scalable variants using empirical Fisher information~\cite{amari1998natural}, block-diagonal curvature~\cite{singh2020woodfisher}, and structured sparsity~\cite{kurtic2022optimal}, enabling efficient use in pruning, quantization, and compression of modern networks~\cite{dong2017learning,wang2019eigendamage,kuznedelev2024cap}.

% We revisit the OBS framework from a new perspective, employing curvature-based compensation for controlled forgetting in personalized unlearning.
% Building on prior block-wise Fisher curvature approximations, we repurpose OBS beyond pruning and compression to construct a verifiable, ZK–compatible unlearning operator that removes provider-specified knowledge while preserving user-specific adaptation and privacy fidelity.

We revisit the OBS framework from a new perspective, employing curvature-based compensation for controlled forgetting in personalized unlearning.
Building on block-wise Fisher curvature approximations, we repurpose OBS beyond pruning and compression into a verifiable, ZK-compatible unlearning operator that removes provider-specified knowledge while preserving user adaptation and privacy fidelity.

\section{System and Problem Formulation}
\label{sec:problem_formulation}

\point{Notation}
Mathematical notations are listed in Appendix~\ref{sec:appendix_notation_table}.

\paragraph{Overview.}
\begin{figure}[t]
    \centering
    \includegraphics[width=0.99\columnwidth]{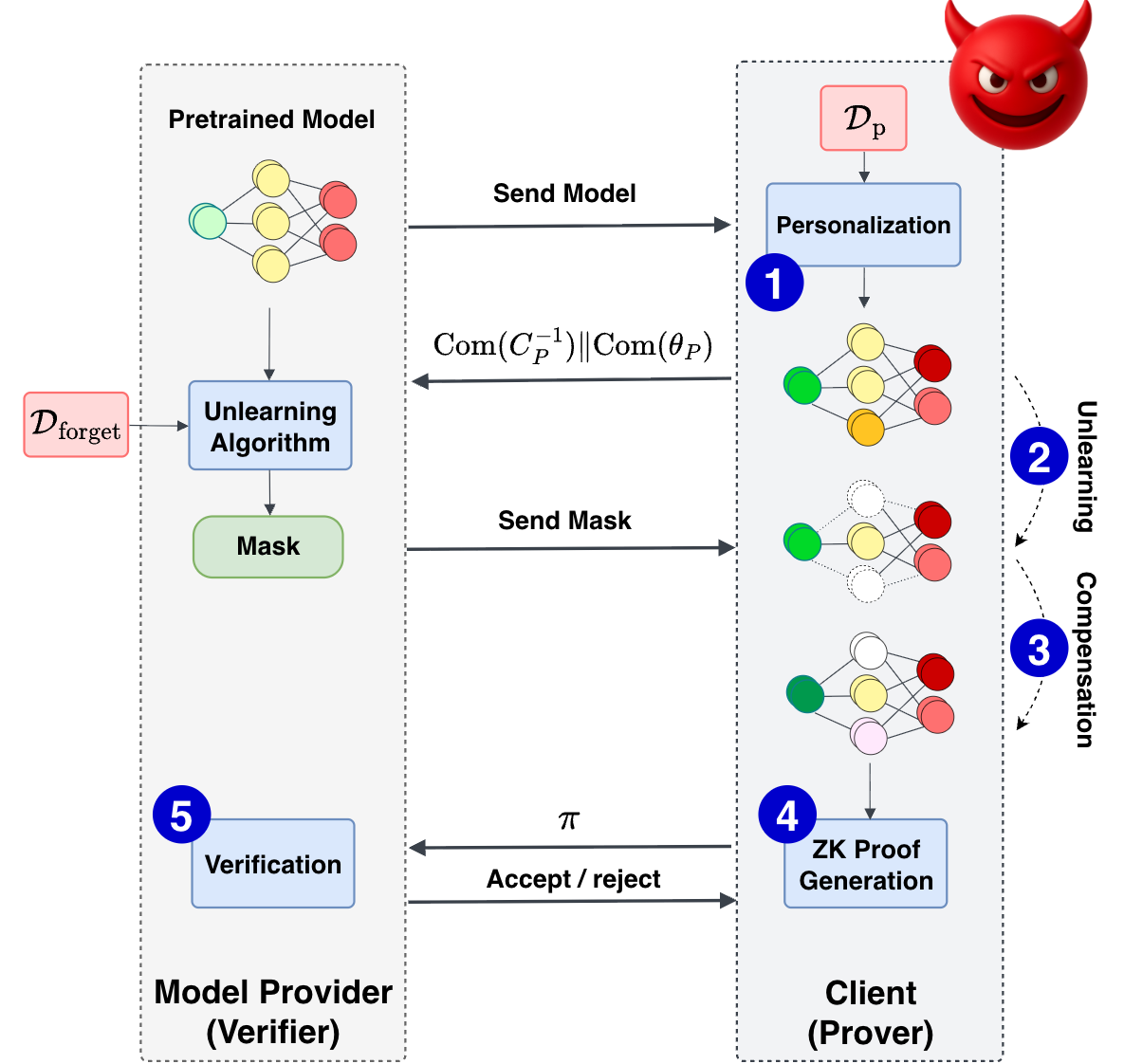}
    \caption{
    Provider sends a public mask removing \forgetset associated weights, the client performs local compensation to recover personalized utility, and ZKP verifies the update’s correctness without exposing the client’s private model.}
    \label{fig:protocol}
\end{figure}

The goal of our methodology is to enable effective and verifiable unlearning for personalized models while ensuring computational efficiency, which aims to minimize time, memory, and proof size overhead during ZKP generation.
Figure \ref{fig:protocol} provides an overview of the proposed framework, illustrating how unlearning is performed and how the corresponding ZK proof certifies its correctness using public and private (committed) inputs.
Section~\ref{sec:problem_formulation} formulates the personalized unlearning problem, defining the setting and objectives.
Section~\ref{section:unlearning_method} presents our proposed unlearning mechanism, which removes the influence of the \forgetset while preserving personalization accuracy through curvature-based masking and compensation.
Finally, Section~\ref{section:proof_generation_method} describes the construction of the ZK-SNARK proof system enabling efficient and confidential verification of the correctness of the unlearning procedure.

\paragraph{Setting.}
Let \(\mathcal{Z}=\mathcal{X}\times\mathcal{Y}\) and let \(\ell(\theta;z)\) be a per-example loss for parameters \(\theta\in\Theta\subseteq\mathbb{R}^d\).
A model provider trains an initial model on pretraining data \(D=\{z_i\}_{i=1}^N\) by empirical risk minimization:
\begin{equation}
\theta_0 \in \arg\min_{\theta\in\Theta}\; L(\theta;D)
\;:=\;
\frac{1}{|D|}\sum_{z\in D}\ell(\theta;z).
\label{eq:erm}
\end{equation}
A client holds a private personalization set \(D_p\) and obtains a personalized model by applying a fixed operator \(P\) (e.g., short-horizon SGD, low-rank adapters such as LoRA~\cite{hu2022lora}, or linear probes): $\theta_p \;=\; P(\theta_0;\, D_p).$

\paragraph{Deletion request and gold standard.}
Upon a deletion request, the provider identifies a \forgetset \(D_f\subseteq D\) and defines the retain set \(D_r:=D\setminus D_f\).
The exact unlearning model, serving as the \emph{gold-standard} of unlearning, retrains on $D_r$ and then personalizes on $D_p$ using the same $P$:
\begin{equation}
\theta_r^\star \in \arg\min_{\theta\in\Theta} L(\theta;D_r),
\qquad
\theta^\star := P(\theta_r^\star;\, D_p).
\label{eq:counterfactual}
\end{equation}
Here, \(\theta^\star\) is the counterfactual model that would have arisen had \(D_f\) never influenced pretraining.

\paragraph{Why approximate unlearning.}
% Exact unlearning in \eqref{eq:counterfactual} is often impractical: (i) the client does not have access to \(D_r\) or \(D_f\); (ii) retraining and re-personalizing per request is computationally prohibitive; (iii) producing zero-knowledge proofs for full retraining and all personalization steps is beyond practical proving budgets.
% We therefore seek a client-side \emph{approximate unlearning} procedure that operates on \(\theta_p\) and aims to match the behavior of \(\theta^\star\).
Exact unlearning in~\eqref{eq:counterfactual} is often impractical for several reasons:
(i) the client does not have access to either $D_r$ or $D_f$;
(ii) retraining and re-personalizing for each deletion request scale with the total number of optimization steps, requiring approximately $\mathcal{O}(E_r |D_r| + E_p |D_p|)$ forward and backward passes, where $E_r$ and $E_p$ denote the number of epochs on $D_r$ and $D_p$, respectively;
and (iii) generating zero-knowledge proofs for full retraining and personalization far exceeds practical proving budgets.
We therefore seek a client-side \emph{approximate unlearning} procedure that operates directly on $\theta_p$, and approximates the behavior of $\theta^\star$.

\paragraph{Prediction-level alignment to the gold standard.}
Let \(p(\cdot\mid x;\theta)\) denote the predictive distribution.
We quantify agreement with the gold standard using a divergence \(d\) between predictive distributions; following common practice in unlearning, we take \(d\) to be the forward Kullback--Leibler divergence,
\[
d\!\left(p(\cdot\mid x;\theta),\,p(\cdot\mid x;\theta^\star)\right)
= \mathrm{KL}\!\left(p(\cdot\mid x;\theta)\,\|\,p(\cdot\mid x;\theta^\star)\right).
\]
We evaluate alignment separately on the personalization and forget domains:
\begin{align}
\mathcal{A}_i(\theta,\theta^\star)
&:= \frac{1}{|D_i|}\sum_{(x,y)\in D_i}
d\!\left(p(\cdot\mid x;\theta),\,p(\cdot\mid x;\theta^\star)\right).
\label{eq:alignments}
\end{align}
Alignment on the personalization domain, \(D_{p}\), preserves personalization; alignment on the forget domain, \(D_f\), enforces that the unlearned model behaves like the retrained model that never used \(D_f\).

\paragraph{Approximate personalized unlearning (definition).}
Given \((\theta_0, D, D_f)\) and a client’s \(D_p\), an approximate unlearning procedure maps \(\theta_p\) to \(\theta_u\) such that, for tolerances \(\varepsilon_p,\varepsilon_f\ge 0\),
\begin{equation}
\mathcal{A}_p(\theta_u,\theta^\star)\;\le\;\varepsilon_p
\qquad\text{and}\qquad
\mathcal{A}_f(\theta_u,\theta^\star)\;\le\;\varepsilon_f.
\label{eq:apu-definition}
\end{equation}

Condition~\eqref{eq:apu-definition} specifies success relative to the gold standard while remaining algorithm-agnostic.
Our method (Section~\ref{section:unlearning_method}) constructs \(\theta_u\) from \(\theta_p\) using only client-side resources and targets small tolerances in \eqref{eq:apu-definition}.
Here, \(\varepsilon_f\) captures forgetting on \(D_f\) in prediction space (the model’s predictions on \(D_f\) align with those of the counterfactual (exact unlearning) that never used \(D_f\)),
whereas \(\varepsilon_p\) guarantees that personalization effectiveness on \(D_p\) is preserved.
% The gold standard \(\theta^\star\) is used only as an (offline) reference for analysis and reporting; it is not required at unlearning time.
The gold standard $\theta^\star$ is used exclusively for offline evaluation and is not computed by either party in deployment: constructing $\theta^\star$ requires access to both $D_r$ (provider side) and $D_p$ (client side). It plays no role in the unlearning protocol or its ZK verification.

\paragraph{Feasible unlearning operators.}
We work with a class \(\mathfrak{U}\) of feasible unlearning operators \(U:\Theta\!\to\!\Theta\) acting on the personalized model: $\theta_u \;=\; U(\theta_p;\,\Psi)$,
% \[
% % \theta_u \;=\; U(\theta_p;\,\Psi,\,\zeta),

% \]
where \(\Psi\) denotes public, provider-agreed traceability artifacts.
% and \(\zeta\) internal randomness.
This section is method-agnostic. In our instantiation, \(\Psi\) is a \emph{sparse mask} (see Section~\ref{section:unlearning_method}).

% \paragraph{Roles and zk verification.}
% % In deployment, the \emph{client (prover)} holds \(D_p\) and \(\theta_p\) and produces \(\theta_u\) by the agreed approximate unlearning transformation; the \emph{provider (verifier)} holds \(\theta_0\), \(D_f\), and \(D_r\).
% % The verification goal (Section~\ref{sec:zk}) is to prove in zero knowledge that \(\theta_u\) was obtained by correctly executing the agreed unlearning transformation on \(\theta_p\), without revealing \(D_p\).
% % This zk certification is orthogonal to the learning objective; alignment in \eqref{eq:alignments} pertains to the machine-learning evaluation of unlearning quality.

% In deployment, the \emph{client (prover)} holds $D_p$ and $\theta_p$, and applies an agreed operator (approximate unlearning) $U \in \mathfrak{U}$ to produce $\theta_u$.
% The \emph{provider (verifier)} holds $\theta_0$, $D_f$, and $D_r$.
% The verification goal (Section~\ref{section:proof_generation_method}) is to prove, in zero knowledge, that $\theta_u$ was obtained by correctly executing the agreed transformation on $\theta_p$, without revealing $D_p$ and $\theta_p$.
% Empirical evaluation of unlearning quality is conducted independently of these roles; zk certification concerns only procedural correctness, whereas alignment in~\eqref{eq:alignments} pertains to the machine-learning evaluation of unlearning quality.

\paragraph{System and Threat Model.}
Our goal is to let the provider (verifier) \emph{cryptographically verify} that the client (prover) applied the agreed approximate unlearning operator \(U\in\mathfrak{U}\) to its personalized model \(\theta_p\), producing 
% \(\theta_u=U(\theta_p;\Psi,\zeta)\),
\(\theta_u=U(\theta_p;\Psi)\),
without learning \(\theta_p\) or any information about \(D_p\).

\emph{Parties and data.}
The \textbf{model provider} holds \(\theta_0\), the pretraining corpus \(D\), the forget/retain split \((D_f,D_r)\), and publishes traceability artifacts \(\Psi\) (e.g., a sparse mask \(m^\star\)).
The \textbf{client} holds private data \(D_p\) and its personalized model \(\theta_p=P(\theta_0;D_p)\), and executes \(U(\cdot;\Psi)\) locally to obtain \(\theta_u\).
As in prior work~\cite{kang2022scaling,weng2023pvcnn,maheri2025telesparse}, the model \emph{architecture} is public; model \emph{weights} (\(\theta_p,\theta_u\)) remain private.

% Public input to the circuit is $\Psi$.
% We assume a binding commitment \(\mathsf{Com}(\theta_p)\) to the client’s personalized model is already available to the provider (e.g., from a separate proof-of-personalization); establishing that commitment is out of scope and can be realized by zero-knowledge training/personalization proofs (e.g., \cite{garg2023experimenting,abbaszadeh2024zero,sun2024zkdl}).

% --- Insert in Sec. 3, right before: \emph{Proof system primitives.} ---

\emph{Workflow (reader map).}
At a high level, the provider publishes a public traceability artifact~$\Psi$ that specifies the agreed unlearning transformation.
Given~$\Psi$, the client locally computes the unlearned model $\theta_u = U(\theta_p;\Psi)$ and generates a zero-knowledge proof $\pi$ certifying operator compliance for the committed inputs and outputs.
The provider then verifies $\pi$ against the public inputs $(\Psi,\Com(\theta_p),\Com(\theta_u))$, thereby certifying correct execution of~$U$ without learning the client’s private model parameters or personalization data.

\emph{Proof system primitives.}
In a ZK setting, verification is conducted through an arithmetic circuit that encodes the agreed computation \(U\).
A concise overview of circuit representations, constraint systems, and polynomial commitment schemes used in ZK-SNARKs is provided in Appendix~\ref{sec:zk_background}.
Here, the client acts as the \emph{prover} and generates a proof~\(\pi\) attesting that the committed personalized model~\(\theta_p\) was correctly transformed into~\(\theta_u\) under the public artifact~\(\Psi\).
The provider acts as the \emph{verifier} and checks the validity of~\(\pi\) without accessing any private inputs.
The public inputs to the circuit are the traceability artifact~\(\Psi\) and the cryptographic commitments~\(\mathsf{Com}(\theta_p)\) and~\(\mathsf{Com}(\theta_u)\), which bind model parameters to their corresponding proofs without revealing them.
% It ensures that the agreed unlearning step was actually applied to the committed model and cannot be silently skipped or altered, which gives the basic trust anchor on which empirical forgetting evaluation then builds.
This guarantee ensures that the agreed unlearning step was actually applied to the committed model and was not silently skipped or altered, thereby providing the basic trust anchor on which empirical forgetting evaluation builds.
Formal definitions of the proof system properties—soundness, correctness, and zero-knowledge—as well as the binding and hiding guarantees of the commitment scheme are provided in Appendix~\ref{sec:appendix_all_properties}.
We assume that a valid commitment~\(\mathsf{Com}(\theta_p)\) has been established beforehand, for example through a proof-of-training or proof-of-personalization protocol~\cite{garg2023experimenting,abbaszadeh2024zero,sun2024zkdl}.

% \emph{Threat model.}
% Adversaries are computationally bounded.
% \textit{(i) Privacy:}
% The proof is zero knowledge (see Appendix~\ref{sec:zk_background}); it reveals no information about \(\theta_p\), \(\theta_u\), or \(D_p\) beyond the public inputs \((\Psi, \mathsf{Com}(\theta_p), \mathsf{Com}(\theta_u))\).
% The model architecture is public, whereas all model weights and client-specific statistics remain part of the private witness.
% \textit{(ii) Integrity:}
% By knowledge soundness, the verifier accepts only if the publicly committed \(\theta_u\) equals the (deterministic) output of the agreed operator \(U\) applied to the committed \(\theta_p\) under \(\Psi\).
% % Any randomness \(\zeta\) is fixed via a public commitment, so the circuit verifies deterministic identities of \(U\).
% After acceptance, all subsequent predictions and updates must reference \(\mathsf{Com}(\theta_u)\), ensuring the deployed model is exactly the certified output of \(U\).
% Operationally, this means the next predictions are produced by the committed \(\theta_u\), not by any model that still encodes \forgetset patterns. 
% All security properties are inherited from Halo2~\cite{zcash-halo2}; formal definitions and assumptions appear in Appendix~\ref{sec:appendix_all_properties}.

\emph{Threat model.}
Adversaries are computationally bounded.
\textit{(i) Privacy (curious verifier/provider):}
The verifier (including a curious provider) observes only the public inputs \((\Psi, \mathsf{Com}(\theta_p), \mathsf{Com}(\theta_u))\) and an argument \(\pi\).
By the zero-knowledge property (Appendix~\ref{sec:zk_background}) and the hiding property of $\mathsf{Com}(\cdot)$, the proof transcript reveals nothing about $\theta_p$, $\theta_u$, or $D_p$ beyond what is already implied by the public inputs.
The model architecture is public, whereas all model weights and client-specific statistics remain part of the private witness.
\textit{(ii) Integrity (dishonest client/prover):}
A dishonest prover may attempt to produce an accepting proof for some \(\theta_u \neq U(\theta_p;\Psi)\).
By knowledge soundness, the verifier accepts only if the publicly committed \(\theta_u\) equals the (deterministic) output of the agreed operator \(U\) applied to the committed \(\theta_p\) under \(\Psi\).
After acceptance, all subsequent predictions and updates must reference \(\mathsf{Com}(\theta_u)\), ensuring the deployed model is exactly the certified output of \(U\).
\textit{(iii) Black-box leakage (unlearning quality):}
An external party may have black-box access to the unlearned model outputs; this is not a protocol break.
Any residual dependence on \(D_f\) is treated as unlearning quality and evaluated separately in our experiments measured by MIA success in Table~\ref{tab:main_results}.
Operationally, this means the next predictions are produced by the committed \(\theta_u\), not by any model that still encodes \forgetset patterns.
All security properties are inherited from Halo2~\cite{zcash-halo2}; formal definitions and assumptions appear in Appendix~\ref{sec:appendix_all_properties}.

\emph{Scope.}
% We do not address secure erasure of the client’s pre-unlearning model (an archival copy of \(\theta_p\) may persist on the device).
We do not address secure erasure of the client’s pre-unlearning model, as an archival copy of \(\theta_p\) may persist on the device. 
Secure deletion is a challenging problem in its own right, often requiring trusted hardware-based method~\cite{hunt2018ryoan,wu2024secgpt}, and is considered orthogonal to our goal of verifiable unlearning.
Our guarantee is \emph{verifiable use}: after acceptance, any future inference or update must reference \(\mathsf{Com}(\theta_u)\), so deployed predictions are produced by the certified unlearned model, and do not rely on parameters that encode \forgetset patterns.
Operational controls (e.g., requiring a valid proof per update/inference) can enforce this usage constraint, but the physical deletion of prior weights is out of scope.

% \paragraph{Procedural ZK guarantee vs.\ semantic unlearning.}
% Our ZK proof certifies \emph{operator compliance}, not semantic or perfect forgetting: an accepting proof attests that the committed unlearned parameters were obtained by applying the agreed deterministic unlearning operator to the committed personalized model under the public artifact, i.e., $\theta_u = U(\theta_p;\Psi)$ for the committed witness.
% By knowledge soundness and zero-knowledge, a computationally bounded prover cannot produce an accepting proof unless this relation holds, while the proof transcript reveals neither $\theta_p$, $\theta_u$, nor any information about $D_p$ beyond the public inputs.
% Since $U$ is an \emph{approximate} substitute for the counterfactual gold-standard model in~\eqref{eq:counterfactual}, any remaining dependence on $D_f$ is treated as \emph{unlearning quality} and is evaluated empirically (e.g., via prediction-level alignment in~\eqref{eq:alignments}), rather than being part of the cryptographic guarantee.
% Empirical unlearning quality is evaluated separately (e.g., via alignment in~\eqref{eq:alignments}); the ZK proof certifies only procedural correctness.

\paragraph{Procedural ZK guarantee vs.\ semantic unlearning.}
An accepting proof certifies \emph{procedural correctness}: the committed output satisfies $\theta_u = U(\theta_p;\Psi)$ for the committed witness, while revealing neither $\theta_p$, $\theta_u$, nor $D_p$ beyond public inputs.
Semantic forgetting is not a cryptographic guarantee; because $U$ is approximate relative to the gold standard in~\eqref{eq:counterfactual}, unlearning quality is assessed empirically (e.g., via prediction-level alignment in~\eqref{eq:alignments}).

% Empirical evaluation of unlearning quality is conducted independently of these roles; zk certification concerns only procedural correctness, whereas alignment in~\eqref{eq:alignments} pertains to the machine-learning evaluation of unlearning quality.

\emph{Design goals for proof efficiency.}
We aim for ZK circuits that (i) minimize the number of constraints by \emph{verifying} operator-specific identities instead of re-running heavy optimization (e.g., SGD) or training inside the circuit; (ii) utilise only \emph{linear} operations (matrix–vector products, inner products), to avoid non-linear activations and large lookup tables that dominate SNARK cost~\cite{maheri2025telesparse}; and (iii) \emph{avoid stochastic optimization inside the circuit}~---~we verify deterministic operator identities and eliminate randomness 
% or fix it via a publicly committed seed
~---~thereby preventing prover adaptivity and the forging attacks observed in verification–unlearning settings that leverage SGD randomness~\cite{DBLP:conf/icml/ZhangCSL24}.

In Section~\ref{section:proof_generation_method}, we instantiate these goals with a block-wise circuit organization and a sparse-mask interface, and we avoid non-linear activations during verification, thereby meeting the above efficiency targets.

% \textcolor{red}{\textbf{Probably need to make this subsection as a separate section then include the threat model (what is goal of MU verification and what is assumptions (commitment on personalized model is available, after unlearning the model still go through same process to commit to data and further personalization steps.))}} $\to$ Needs to exclude the attacker abilities/goal of "Verification of Machine Unlearning is Fragile" paper and any forging attack (save \forgetset information). 

% \textcolor{red}{Stress on the fact that goal of verification is that next prediction of model by new committed model does not include the pattern in the \forgetset}

\section{Proposed Approximate Unlearning Algorithm}
\label{section:unlearning_method}

\paragraph{From formulation to method.}
Consistent with Definition~\eqref{eq:apu-definition}, we instantiate a sparse, client-side operator \(U\in\mathfrak{U}\) that acts directly on the personalized model to increase loss on \(D_f\) while preserving utility on \(D_p\).
We work with the decomposition
\begin{equation}
\theta_p=\theta_0+\Delta_p,\qquad \Delta_p=BA,\qquad \mathrm{rank}(\Delta_p)\le r\ll d,
\label{eq:theta-p-decomp}
\end{equation}
which captures common adapter-style personalization (e.g., LoRA). The algorithm itself does not assume low rank.

\paragraph{Saliency and mask selection \emph{at the personalized model}.}
% Let \(\Delta L_f := L(\theta_p+\delta w;D_f)-L(\theta_p;D_f)\) denote the change of the \forgetset loss under a parameter perturbation \(\delta w\). 
We denote by \(\Delta L_f\) the change in the \forgetset loss when the personalized parameters are perturbed by \(\delta w\), as shown below:
\begin{equation}
\Delta L_f := L(\theta_p + \delta w; D_f) - L(\theta_p; D_f).
\end{equation}
A successful unlearning operation should \emph{maximize} \(\Delta L_f\), i.e., increase the loss on \(D_f\), reflecting stronger forgetting.
Assuming \(L(\cdot;D_f)\) is \(c^2\) with locally Lipschitz Hessian near \(\theta_p\), a second-order expansion gives
\begin{align}
\Delta L_f
&=
g_f(\theta_p)^\top\delta w
+\tfrac{1}{2}\,\delta w^\top H_f(\theta_p)\,\delta w
+R_3(\delta w),
\nonumber\\
&\quad
\|R_3(\delta w)\|\le c\|\delta w\|^3.
\label{eq:taylor-second}
\end{align}
where \(g_f(\theta)=\nabla_\theta L(\theta;D_f)\) and \(H_f(\theta)=\nabla^2_\theta L(\theta;D_f)\).
Zeroing coordinate \(i\) at \(\theta_p\) corresponds to \(\delta w=-\theta_{p,i}e_i\), yielding the local forgetting gain
\begin{equation}
\Delta L_f^{(i)}
\approx
-\,g_{f,i}(\theta_p)\,\theta_{p,i}
+\tfrac{1}{2}\,H_{f,ii}(\theta_p)\,\theta_{p,i}^{2}.
\label{eq:delta-single}
\end{equation}
Coordinates that yield larger \(\Delta L_f^{(i)}\) contribute more to the desired loss increase and are therefore more suitable for masking.
We rank parameters by the per-coordinate saliency
\[
S_i(\theta_p;D_f)
:= -\,g_{f,i}(\theta_p)\,\theta_{p,i}
+\tfrac{1}{2}\,[C_f(\theta_p)]_{ii}\,\theta_{p,i}^{2},
\]
where \(C_f(\theta_p)\simeq \mathrm{diag}(H_f(\theta_p))\) is a diagonal curvature proxy (Hessian-diag or diagonal empirical Fisher) with damping.
The top-\(k\) indices that maximize the cumulative predicted forgetting gain are selected:
\begin{equation}
m^\star\in
\arg\max_{m\in\{0,1\}^d}\;\sum_{i=1}^{d} m_i\,S_i(\theta_p;D_f)
\quad \text{s.t.} \quad \|m\|_0=k,
\label{eq:mask-select-theta-p}
\end{equation}
and we denote \(M:=\mathrm{supp}(m^\star)\), \(C:=[d]\setminus M\).

\paragraph{Masking the personalized model and loss decomposition.}
Applying the mask to the personalized parameters,
\begin{equation}
\theta_u \;=\; \theta_p + \delta w_m + \delta w_c,
\quad
\delta w_m:= -\,\theta_p\odot m^\star,
\quad
m^\star\odot \delta w_c=\mathbf{0}.
\label{eq:masked-theta}
\end{equation}
Inserting \(\delta w=\delta w_m+\delta w_c\) into \eqref{eq:taylor-second} and partitioning by \((M,C)\) yields

\begin{align}
\Delta L_f
&\approx
\underbrace{\sum_{i\in M}
\Big(-\,g_{f,i}(\theta_p)\,\theta_{p,i}
+\tfrac{1}{2}\,H_{f,ii}(\theta_p)\,\theta_{p,i}^2\Big)}_{\text{mask-only increase}}
\nonumber\\[3pt]
&\quad+\;
\underbrace{g_{f,C}(\theta_p)^\top\delta w_c}_{\text{residual linear on }C}
\;+\;
\underbrace{\delta w_c^\top[H_f(\theta_p)]_{C,M}\,\delta w_m^M}_{\text{cross curvature}}
\nonumber\\[3pt]
&\quad+\;
\underbrace{\tfrac{1}{2}\,\delta w_c^\top[H_f(\theta_p)]_{C,C}\,\delta w_c}_{\text{quadratic on }C}
\;+\;R_3.
\label{eq:forget-decomp}
\end{align}

% \paragraph{Why do the cross and compensation terms remain controlled?}
% While the compensation step helps maintain personalization utility, it is not intended to undo the forgetting effect—our goal is to keep the overall change \(\Delta L_f\) positive and large. 
% In \eqref{eq:forget-decomp}, the non-mask part on \(C\) has three pieces: a residual linear term (from \(g_{f,C}\)), a cross-curvature term (from \([H_f]_{C,M}\)), and a quadratic term on \(C\).
% (i) \emph{Residual gradient on \(C\).} The mask \eqref{eq:mask-select-theta-p} concentrates \(D_f\)-sensitivity on \(M\), keeping the spillover \(g_{f,C}(\theta_p)\) small.
% (ii) \emph{Cross-block curvature.} We partition by architectural modules (layers/heads) for which the \(D_f\) Hessian is empirically near block-diagonal; hence \([H_f(\theta_p)]_{C,M}\) is moderate and the cross term is limited.
% (iii) \emph{Quadratic dominance on \(C\).} We apply standard damping to the \(C\)-block, making it positive-definite and penalizing large compensations, which caps how negative the non-mask contribution can be.
% A concise derivation with explicit bounds for these non-mask terms is provided in Appendix~\ref{app:cross-comp-analysis}.

\paragraph{Why do the cross and compensation terms remain controlled?}
While the compensation step helps maintain personalization utility, it is not intended to undo the forgetting effect; our goal is to keep the overall change \(\Delta L_f\) positive and large. 
In \eqref{eq:forget-decomp}, the non-mask contribution on \(C\) consists of three components: a residual linear term arising from \(g_{f,C}\), a cross-curvature term from \([H_f]_{C,M}\), and a quadratic term defined over \(C\). 
The residual gradient term remains small because the mask in \eqref{eq:mask-select-theta-p} concentrates the \(D_f\)-sensitivity on the masked subset \(M\), leaving minimal spillover to \(C\). 
The cross-curvature component is moderate, as the model is partitioned by architectural modules (layers or heads) for which the \(D_f\) Hessian is empirically close to block-diagonal, thereby limiting \([H_f(\theta_p)]_{C,M}\). 
Finally, the quadratic term is controlled through standard damping applied to the \(C\)-block, which ensures positive definiteness and penalizes large compensations, bounding the magnitude of any negative contribution. 
A detailed derivation with explicit bounds for these non-mask terms is provided in Appendix~\ref{app:cross-comp-analysis}.

% \paragraph{OBS compensation on \(C\) anchored at the personalized model.}
% To protect personalization on \(D_p\), we enforce \((\theta_u)_M=0\) while minimizing a quadratic surrogate of \(L(\cdot;D_p)\) at \(\theta_p\), assuming approximate stationarity:
% \begin{equation}
% g_p(\theta_p)=\nabla_\theta L(\theta_p;D_p)\approx \mathbf{0}.
% \label{eq:stationary}
% \end{equation}
% Let \(C_p:=F_p(\theta_p)+\lambda I\) be a damped empirical Fisher on \(D_p\),
% \[
% F_p(\theta_p)=\frac{1}{|D_p|}\sum_{(x,y)\in D_p}\nabla_\theta \ell(\theta_p;(x,y))\,\nabla_\theta \ell(\theta_p;(x,y))^\top,\qquad \lambda>0.
% \]
% We solve the group-OBS quadratic program
% \begin{equation}
% \min_{\delta w}\ \tfrac{1}{2}\delta w^\top C_p\,\delta w
% \quad\text{s.t.}\quad
% E_M^\top \delta w + w_{p,M} = 0,\qquad w_p:=\theta_p,\ E_M=[e_i]_{i\in M},
% \label{eq:obs-qp}
% \end{equation}
% whose KKT solution is
% \begin{equation}
% \boxed{
% \delta w^\star
% = -\,C_p^{-1}\,E_M\,\big(E_M^\top C_p^{-1}E_M\big)^{-1}\, w_{p,M}.
% }
% \label{eq:obs-full}
% \end{equation}
% The final parameters \(\theta_u=\theta_p+\delta w^\star\) obey \((\theta_u)_M=0\).
% In practice, we compute \(C_p^{-1}E_M\) via damped Fisher-vector products and conjugate gradients, and invert only the \(|M|\times|M|\) Schur complement.

\paragraph{OBS compensation on \(C\) anchored at the personalized model.}
While pruning yields a binary mask that removes parameters to induce forgetting, directly applying such a mask to a personalized model is problematic. 
Personalized models have adapted their parameters to sensitive local data; thus, indiscriminate pruning based on global importance scores can erase features essential for the personalization task, leading to significant accuracy degradation. 
This degradation arises because pruning typically targets neurons influential to the \forgetset without accounting for their contribution to the retained personalized data, as neural representations often entangle multiple data sources. 
To mitigate this effect, we leverage the Optimal Brain Surgeon (OBS) framework~\cite{lecun1989optimal,hassibi1993optimal,kuznedelev2023cap}, which uses second-order information to optimally adjust the remaining weights after pruning.

To protect personalization on \(D_p\), we enforce \((\theta_u)_M=0\) while minimizing a quadratic surrogate of \(L(\cdot;D_p)\) at \(\theta_p\), assuming approximate stationarity:
\begin{equation}
g_p(\theta_p)=\nabla_\theta L(\theta_p;D_p)\approx \mathbf{0}.
\label{eq:stationary}
\end{equation}
Let \(C_p:=F_p(\theta_p)+\lambda I\) be a damped empirical Fisher on \(D_p\) such that $\lambda>0$,
\[
F_p(\theta_p)=\frac{1}{|D_p|}\sum_{(x,y)\in D_p}\nabla_\theta \ell(\theta_p;(x,y))\,\nabla_\theta \ell(\theta_p;(x,y))^\top.
\]

We solve the group-OBS quadratic program
\begin{equation}
\begin{aligned}
\min_{\delta w}\;&\tfrac{1}{2}\delta w^\top C_p\,\delta w
\quad\text{s.t.}\quad
E_M^\top \delta w + w_{p,M} = 0,\\[3pt]
&w_p:=\theta_p,\qquad E_M=[e_i]_{i\in M},
\end{aligned}
\label{eq:obs-qp}
\end{equation}

whose KKT solution is

\begin{equation}
\boxed{
\delta w^\star
= -\,C_p^{-1}\,E_M\,\big(E_M^\top C_p^{-1}E_M\big)^{-1}\, w_{p,M}.
}
\label{eq:obs-full}
\end{equation}

The final parameters \(\theta_u=\theta_p+\delta w^\star\) obey \((\theta_u)_M=0\).
In practice, we compute \(C_p^{-1}E_M\) via damped Fisher-vector products and conjugate gradients, and invert only the small \(|M|\times|M|\) Schur complement. 
Following prior work~\cite{kurtic2022optimal,kuznedelev2023cap}, we adopt a block-wise Fisher structure for computational efficiency while accurately capturing the user-specific curvature.

\paragraph{Provider-side mask selection at \(\theta_0\) (efficiency, privacy, and traceability).}
Selecting \(m^\star\) at \(\theta_p\) via \eqref{eq:mask-select-theta-p} is client-specific and, in deployment, misaligned with the privacy constraint. The provider (who owns \(D_f\)) cannot evaluate \(g_f(\theta_p)\) without access to \(\theta_p\), and the client (who owns \(\theta_p\)) cannot evaluate \(g_f\) without access to \(D_f\). We therefore compute the mask once \emph{provider-side} at the pretrained weights,
\begin{equation}
S_i(\theta_0;D_f)
:= -\,g_{f,i}(\theta_0)\,\theta_{0,i}
+\tfrac{1}{2}\,[C_f(\theta_0)]_{ii}\,\theta_{0,i}^{2},
\label{eq:snip-saliency-theta0}
\end{equation}
and publish the resulting binary support \(M=\mathrm{supp}(m^\star)\) as the traceability artifact \(\Psi\).
% This choice is also computationally attractive (one mask for all clients) and ZK-friendly (the heavy proving burden for mask selection stays provider-side; see Section~\ref{section:proof_generation_method}).
This choice is computationally efficient (one mask for all clients) and ZK friendly, as the proving burden for mask selection remains provider side (see Section~\ref{section:proof_generation_method}).
\medskip
% \noindent\textbf{Stability justification.}
The provider-side score in \eqref{eq:snip-saliency-theta0} closely approximates the client-side score at \(\theta_p\).
Write \(\theta_p=\theta_0+BA\). Under the \(C^3\) smoothness assumed above and using Taylor expansions at \(\theta_0\),
\begin{align}
g_f(\theta_0\!+\!BA)
&= g_f(\theta_0) + H_f(\theta_0)\,BA + R_g, \label{eq:gf-expand}\\[2pt]
H_f(\theta_0\!+\!BA)
&= H_f(\theta_0) + \mathcal{T}_f(\theta_0)[BA] + R_H, \label{eq:Hf-expand}\\[2pt]
\|R_g\| = O(&\|BA\|^2), \quad \|R_H\| = O(\|BA\|^2). \nonumber
\end{align}
which implies a first-order perturbation of the per-coordinate saliency:
\begin{equation}
\begin{aligned}
\big|S_i(\theta_p;D_f)-S_i(\theta_0;D_f)\big|
\le \alpha_i\,\|BA\| + O(\|BA\|^2),\\[3pt]
% &\alpha_i = O\!\Big(\,|{\theta_0}_i|\,\|H_f(\theta_0)\|
% + {\theta_0}_i^2\,\|\mathcal{T}_f(\theta_0)\|
% + |c_i(\theta_0){\theta_0}_i - g_{f,i}(\theta_0)|\,\Big).
\alpha_i = O~\!\big(|{\theta_0}_i|\,\|H_f(\theta_0)\| + {\theta_0}_i^2\,\|\mathcal{T}_f(\theta_0)\| + |g_{f,i}(\theta_0)|\big).
\end{aligned}
\label{eq:saliency-stability}
\end{equation}

% where \(c_i(\theta)\) is the damped diagonal curvature proxy used in \eqref{eq:mask-select-theta-p}.
% For adapter-style personalization (low-rank, short-horizon updates), \(\|BA\|\) is small and concentrated in a few modules, so the discrepancy in \eqref{eq:saliency-stability} remains modest in practice.

% \medskip
% % \noindent\textbf{Empirical observation.}
% Across all datasets and personalization regimes evaluated, computing the mask at \(\theta_p\) yields only marginal improvements in forgetting compared to provider-side \(\theta_0\) masks, while incurring a large additional cost for the client (and, in ZK settings, a substantial increase in proof generation due to per-client gradient/curvature evaluation at \(\theta_p\); see Section~\ref{section:proof_generation_method}). Consequently, we adopt \(\theta_0\)-based mask selection by default and use \eqref{eq:saliency-stability} to justify its closeness to the client-specific alternative.

where $c_i(\theta)$ is the damped diagonal curvature proxy used in Equation \ref{eq:mask-select-theta-p}.
For adapter style personalization (low rank, short horizon updates), $\|BA\|$ is small and concentrated in a few modules, so the discrepancy in \eqref{eq:saliency-stability} remains modest in practice. To reduce client cost, and in ZK settings to avoid the substantial increase in proof generation due to per client gradient and curvature evaluation at $\theta_p$ (see Section~\ref{section:proof_generation_method}), we adopt $\theta_0$ based mask selection by default and use \eqref{eq:saliency-stability} to justify its closeness to the client specific alternative.

\paragraph{Operator and implementation.}
The resulting operator is
\begin{equation}
U(\theta_p;\,m^\star)
\;=\;
(\mathbf 1-m^\star)\odot \theta_p \;+\; \delta w^\star,
\label{eq:final-operator}
\end{equation}
where \(m^\star\) is selected once by the provider using \eqref{eq:snip-saliency-theta0}–\eqref{eq:mask-select-theta-p} (approximated at \(\theta_0\)) and \(\delta w^\star\) is computed on the client via \eqref{eq:obs-qp}–\eqref{eq:obs-full}.
This separation allows a single public, provider-agreed mask (traceability artifact \(\Psi\)) and private, client-side compensation, and is ZK-friendly.
As per Section~\ref{sec:problem_formulation}, the counterfactual \(\theta^\star\) is \emph{not} needed to run the algorithm and is used solely for offline evaluation of alignment.

\section{Efficient Zero-Knowledge Proof Generation}
\label{section:proof_generation_method}

\begin{figure}[t]
    \centering
    \includegraphics[width=0.80\columnwidth]{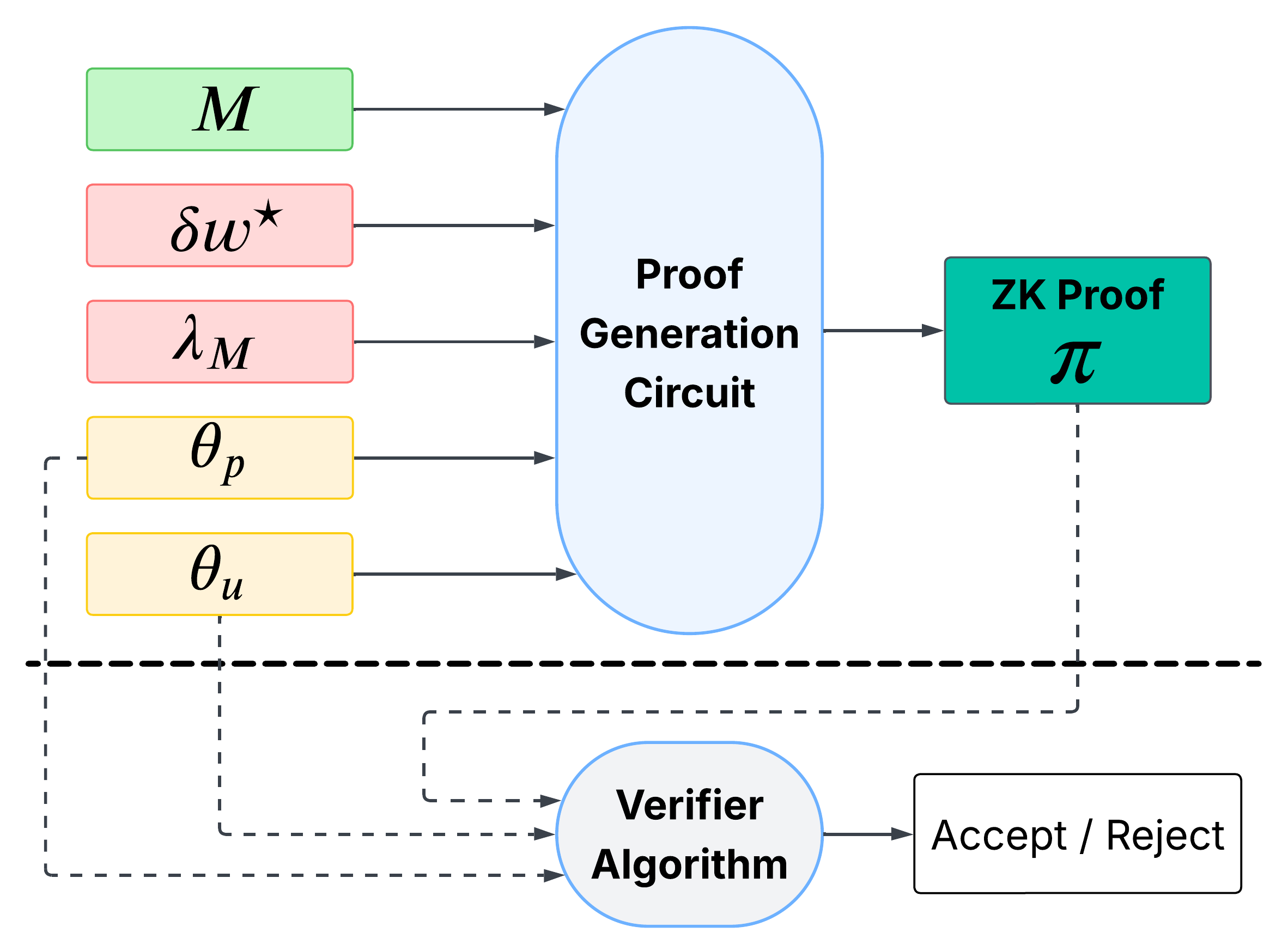}
    \caption{
    Structure of the ZK-SNARK circuit used for verification. 
    The circuit enforces linear constraints corresponding to the unlearning operator.}
    \label{fig:zk_circuit}
\end{figure}

\paragraph{Goal.}
We construct a ZK-SNARK circuit that enables the model provider (verifier) to confirm that the client (prover) correctly applied the agreed unlearning operator \(U\) to its personalized model \(\theta_p\), producing the unlearned model \(\theta_u\), as defined in Section~\ref{section:unlearning_method}, without revealing \(\theta_p\), \(\theta_u\), or any data-dependent quantities from \(D_p\).
The end-to-end prover–verifier workflow corresponding to this circuit is illustrated in Figure~\ref{fig:zk_circuit} and summarized in Algorithm~\ref{alg:verifiable_unlearning}.

\paragraph{Public and private objects.}
Public inputs to the circuit are the traceability artifact \(\Psi\) (mask \(m^\star\) and its support \(M\)), together with the commitments
\(\mathsf{Com}(\theta_p)\), \(\mathsf{Com}(\theta_u)\), and \(\mathsf{Com}(C_p)\).
Private witnesses include \(\theta_p,\theta_u\), the compensation vector \(\delta w\) (aggregating \(\delta w_m+\delta w_c\)), the Lagrange multipliers \(\lambda_M\), and the block-wise Fisher proxy \(C_p\) (opened privately to match its public commitment).
As stated in Section~\ref{sec:problem_formulation}, \(C_p\) is computed \emph{offline} on the client using a small subsample of \(D_p\) (e.g., 1K samples). It relies solely on first-order gradients (empirical Fisher approximation), requires no second-order differentiation, and does not need to be recomputed per unlearning request.

\paragraph{Circuit verification of the Group-OBS certificates.}
The circuit enforces the algebraic equalities implied by the Group-OBS KKT system (Section~\ref{section:unlearning_method}), together with model assembly and mask constraints:
\begin{subequations}
\begin{alignat}{2}
\textbf{(Assembly)} \;& \theta_u = \theta_p + \delta w, \label{eq:zk-assembly} \\[4pt]
\textbf{(Mask feasibility)} \;& E_M^\top \delta w + w_{p,M} = 0, \\[2pt]
& w_{p,M} := E_M^\top \theta_p, \label{eq:zk-feas} \\[4pt]
\textbf{(KKT stationarity)} \;& C_p\,\delta w + E_M\,\lambda_M = 0. \label{eq:zk-stat}
\end{alignat}
\end{subequations}
Here \(E_M \in \mathbb{R}^{d\times k}\) stacks the standard basis vectors for the masked coordinates (\(|M|=k\)).
Eq.~\eqref{eq:zk-feas} ensures that the masked parameters are zeroed (\((\theta_u)_M=0\)),
while Eq.~\eqref{eq:zk-stat} verifies that \(\delta w\) satisfies the first-order optimality (KKT) condition of the Group-OBS program.
Because the objective is strictly convex (\(C_p \succ 0\)) with linear constraints,
Eqs.~\eqref{eq:zk-feas}--\eqref{eq:zk-stat} are \emph{necessary and sufficient} for optimality—any feasible pair \((\delta w,\lambda_M)\) satisfying them must correspond to the unique primal solution \(\delta w^\star\) of the OBS system.

\paragraph{Block-wise Fisher and linear algebraic checks.}
We adopt the block decomposition of Section~\ref{section:unlearning_method}, writing
We partition the personalized curvature matrix \(C_p\) into \(B\) disjoint blocks,
\[
C_p = \operatorname{diag}(C_p^{(1)},\ldots,C_p^{(B)}),
\quad C_p^{(b)} \in \mathbb{R}^{d_b\times d_b}, \ \sum_{b=1}^B d_b = d,
\]
so that each block acts on its corresponding parameter slice \(\delta w^{(b)}\in\mathbb{R}^{d_b}\).
Each block \(C_p^{(b)} \in \mathbb{R}^{d_b\times d_b}\) acts on its corresponding parameter slice \(\delta w^{(b)} \in \mathbb{R}^{d_b}\), producing
\[
y^{(b)} = C_p^{(b)}\,\delta w^{(b)}, \quad y = \bigoplus_{b} y^{(b)}.
\]
The circuit verifies the global condition \(y + E_M \lambda_M = 0\).
All constraints are purely linear-algebraic—matrix–vector products, inner products, and additions—avoiding any nonlinear activations or lookup tables, in line with the efficiency design goals defined in Section~\ref{sec:problem_formulation}.

% \paragraph{Circuit completeness and efficiency.}
% (i) Eq.~\eqref{eq:zk-assembly} binds the public commitments of \(\theta_p\) and \(\theta_u\);  
% (ii) Eq.~\eqref{eq:zk-feas} confirms correct mask application;  
% (iii) Eq.~\eqref{eq:zk-stat} certifies optimality of the compensation vector under the private curvature \(C_p\).  
% Together, these constraints guarantee that the committed output \(\theta_u\) equals the deterministic operator output \(U(\theta_p;\Psi)\) (Eqns.~\eqref{eq:masked-theta}, \eqref{eq:obs-full}), without executing any iterative solver inside the circuit.

\paragraph{Circuit completeness and efficiency.}
Equation~\eqref{eq:zk-assembly} binds the public commitments of \(\theta_p\) and \(\theta_u\), ensuring consistency between the personalized and unlearned model parameters. 
Equation~\eqref{eq:zk-feas} verifies that the unlearning mask has been applied correctly, while Equation~\eqref{eq:zk-stat} certifies the optimality of the compensation vector under the private curvature matrix \(C_p\). 
Together, these constraints guarantee that the committed output \(\theta_u\) exactly matches the deterministic operator output \(U(\theta_p;\Psi)\), as defined in Equations~\eqref{eq:masked-theta} and~\eqref{eq:obs-full}, without executing any iterative solver within the circuit.
An analytical estimate of dimensionality and computational cost is given in Appendix~\ref{sec:appendix_dim_cost}.

\section{Evaluation}
% Our experimental evaluation aims to assess the effectiveness of the proposed unlearning algorithm in personalized settings. Specifically, we focus on: (i) evaluating the extent to which the unlearning algorithm removes information pertaining to the \forgetclass on the personalized model by measuring the model's accuracy on that class; (ii) quantifying the impact of directly applying the unlearning mask on the personalized model's performance; and (iii) determining the degree to which the weight adjustment compensates for any performance degradation resulting from the unlearning process.

\subsection{Experimental Questions}
\label{sec:eqs}

Our evaluation focuses on four key questions.  
\textbf{(EQ1) Forgetting efficacy:} Does \name effectively remove the influence of the designated \forgetset while maintaining model stability?  
\textbf{(EQ2) Personalization retention:} To what extent does the unlearned personalized model preserve its performance on the client’s personalization data?  
\textbf{(EQ3) Verification efficiency:} What level of computation and resources are required to generate and verify the ZKP of unlearning?  
\textbf{(EQ4) Design sensitivity:} How do key algorithmic and structural parameters influence the trade-off between forgetting, retention, and verification cost?  
% \textbf{(EQ5) Robustness:} Do the unlearning and personalization effects generalize across domains, such as transferring from ImageNet to ImageNet-Sketch?
Collectively, they assess the framework’s effectiveness, efficiency, and generality.

% \begin{table*}[!ht]
% \centering
% \caption{Impact of Unlearning and Weight Adjustment on Personalized Model Performance}
% \begin{tabular}{lcc}
% \toprule
% \textbf{Method} & \textbf{Forget Class Accuracy (\%) $\downarrow$} & \textbf{Personalized Accuracy (\%) $\uparrow$} \\
% \midrule
% Personalized Model (Baseline) & 93.7 & 71.5 \\
% After Applying Unlearning Mask & 60.5 & 69.4 \\
% After Weight Adjustment (\textbf{Our method}) & 59.5 & 70.9 \\
% \midrule
% \textit{Improvement over Naive Mask (\%)} & \textbf{1.7} & \textbf{71.4} \\
% \bottomrule
% \end{tabular}
% \label{table:performance}
% \end{table*}

\subsection{Baselines}
\label{sec:baselines}

We evaluate our method against representative approximate and exact unlearning approaches.  

% \textbf{Approximate unlearning.}  
% We consider two gradient-based baselines commonly used in prior work.  
% (i) \emph{Gradient Ascent (GA):}~\cite{graves2021amnesiac,thudi2022unrolling} performs ascent on the \forgetset to increase its loss while ignoring the effect on retained (or personalized data).  
% (ii) \emph{SCRUB}~\cite{kurmanji2023towards}: alternates between gradient ascent on the \forgetset and descent on the retain set, aiming to balance forgetting with utility preservation.  
% Both approaches require iterative optimization, making them costly to verify within zero-knowledge circuits.

% \textbf{Exact unlearning.}  
% As an ideal reference, we include a retrain-on-retain baseline where the model provider retrains the global model on the retain set before redistributing it to clients, who must then re-personalize using their local data.  
% This procedure achieves exact unlearning but is computationally prohibitive, particularly for edge scenarios, and forms the basis of prior ZK-verifiable unlearning schemes~\cite{eisenhofer2025verifiable}.

% These baselines span gradient-based approximate methods to fully retrained models, allowing direct comparison against our zero-shot, ZK-efficient personalized unlearning framework.

\textbf{Approximate unlearning.}  
We evaluate two gradient-based baselines.  
(i) \emph{Gradient Ascent (GA)}~\cite{graves2021amnesiac,thudi2022unrolling} increases the loss on the \forgetset without accounting for retained data.  
(ii) \emph{SCRUB}~\cite{kurmanji2023towards} alternates ascent on the \forgetset and descent on the retain set to balance forgetting and utility.  
Both require iterative optimization, making them expensive to verify in zero-knowledge circuits.

\textbf{Exact unlearning.}  
As a reference, we retrain the model on the retain set and then re-personalize on each client, following~\cite{eisenhofer2025verifiable}.  
This achieves exact unlearning but is computationally infeasible for edge deployment.

These baselines, spanning gradient-based to fully retrained methods, highlight the efficiency advantage of our zero-shot, ZK-verifiable personalized unlearning framework.

\subsection{Experimental Setup}
\label{sec:exp_setup}

\begin{table*}[!t]
\centering
\small
\caption{\textbf{Unlearning efficacy and ZK proof efficiency.}
Left (unlearning): lower is better for Forget Acc.\ and MIA AUC; higher is better for Personal Acc.\ (all in \%).
Right (ZKP): lower is better for proving time (h), peak memory (GB), proof size (MB), and verification time (min).}
\label{tab:main_results}
\vspace{0.4em}
\resizebox{\textwidth}{!}{
\begin{tabular}{lcccccccc}
\toprule
\multirow{2}{*}{\textbf{Method}} &
\multicolumn{3}{c}{\textbf{Unlearning Efficacy}} & &
\multicolumn{4}{c}{\textbf{ZK Proof Efficiency}} \\
\cmidrule(lr){2-4} \cmidrule(lr){6-9}
& \textbf{Forget Acc.} & \textbf{Personal Acc.} & \textbf{MIA AUC} & &
\textbf{Proving Time} & \textbf{Peak Mem.} & \textbf{Proof Size} & \textbf{Verification Time} \\
\midrule
Pre-unlearning model & 89.5 & 81.0 & 54.0 & & -- & -- & -- & -- \\
\midrule
Mask only (4\%) & 50.2~$\pm$~0.1 & 74.3~$\pm$~0.3 & 50.4~$\pm$~0.2 & & 1.5~$\pm$0.1 & 0.5~$\pm$~0.01 & 391.4~$\pm$~0.06 & $9.3$ \\
GA (1 epoch) & 50.2~$\pm$~0.1 & 72.6~$\pm$~0.4 & 50.3~$\pm$~0.1 & & $4.9*10^6$ & 245.3~$\pm$~10.3 & $5.8*10^7$ & $216.7$ \\
SCRUB (1 epoch) & 53.3~$\pm$~0.8 & 79.6~$\pm$~0.4 & 50.6~$\pm$~0.3 & & $9.7*10^6$ & 245.3~$\pm$~10.3 & $1.1*10^8$ & $433.3$ \\
\textbf{\name} & \textbf{50.2}~$\pm$~0.1 & \textbf{80.9}~$\pm$~0.2 & \textbf{50.2}~$\pm$~0.1 & & \textbf{2.0}~$\pm$~0.1 & \textbf{0.7}~$\pm$~0.01 & \textbf{401.5}~$\pm$~1.4 & \textbf{10.0} \\
\midrule
\textit{Exact unlearning} & 50.1 & 81.0 & 50.0 & & $2.5*10^7$ & 245.3~$\pm$~10.3 & $2.98 * 10^8$ & $4.0 * 10^6$ \\
\bottomrule
\end{tabular}
}
\end{table*}

\paragraph{Models and datasets.}
We evaluate two models: Google’s ViT-B/16 for classification and Meta’s OPT-125M for language modeling.
Each model was fine-tuned on a separate personalization set to emulate the provider--client setup.
ViT, pretrained on ImageNet, was personalized on ImageNet-Sketch using full fine-tuning; in Appendix~\ref{sec:appendix-adaptformer}, we additionally evaluate AdaptFormer~\cite{chen2022adaptformer} to demonstrate the extensibility of our method across different personalization algorithms.
For OPT-125M, we first fine-tune on Scala, C, C++, and Java using LoRA~\cite{hu2022lora} to narrow its broad pretraining scope and enable a distinct \forgetset, and then personalize on Rust from CodeParrot’s GitHub Clean dataset, again using LoRA~\cite{hu2022lora}.
Both setups capture domain shifts typical of client-side personalization.

\paragraph{Unlearning and compensation.}
% A \forgetset, \(D_f\), consisting of 33,600 samples, or 2.6\% of the original data, was selected for the ViT. For the LLM, \(D_f\) was chosen to be a subset of the Scala training examples, representing 4.8\% of the original Scala train set and 1.2\% of the overall train set.
A \forgetset, $D_f$, consisting of 33{,}600 samples (2.6\% of the original training data) was selected for the ViT.
For the LLM, $D_f$ was chosen as a subset of the Scala training examples, representing 4.8\% of the Scala training split and 1.2\% of the overall training data.
Unless stated otherwise, we construct $D_f$ by uniform random sampling from the designated source split.
To study a structured (class-conditional) forgetting regime, we additionally evaluate cases where $D_f$ is concentrated in a single (or a small number of) classes; see Appendix~\ref{sec:appendix-classUnl}.

In both cases an unlearning mask is computed on the pretrained model, targeting only the MLP sublayers across all transformer blocks and pruning \(4\%\) of their parameters while leaving attention heads intact, following~\cite{pochinkov2024dissecting,meng2022locating} which show MLP pruning to be more effective for unlearning.  
The saliency score in Section~\ref{section:unlearning_method} is used with curvature estimated from a diagonal block-wise empirical Fisher 
% computed on \(1\%\) of the personalization data.  
All hyperparameters are tuned on a separate validation split, and sensitivity analyses are reported in Section~\ref{sec:ablations}

Appendix~\ref{sec:appendix-forgetsize} further studies sensitivity to the \forgetset size $|D_f|$ by increasing $|D_f|$ beyond our default setting and tracing the resulting forgetting--personalization accuracy trade-off.
In particular, we report how forgetting effectiveness and retention effect as $|D_f|$ grows, providing a clearer picture of robustness under larger unlearning requests.

\paragraph{Evaluation.}
% For the image classification task, accuracy was used as the primary metric to evaluate performance. Membership inference (MIA)~\cite{kurmanji2023towards,maheri2025teleportation} leakage was also evaluated For the generation task, both Top-1 accuracy and perplexity are used.

% For the image-classification task, we use accuracy as the primary utility metric. We additionally evaluate privacy leakage via a membership inference attack (MIA) adapted to unlearning~\cite{kurmanji2023towards}. Following their protocol, we form a binary attack dataset using the unlearned model’s per-example output loss on \emph{forget-set} samples (members) versus an identically-distributed held-out \emph{test} set (non-members), train an attack classifier on a class-balanced split, and then report its performance on a disjoint, balanced held-out evaluation split using the \emph{area under the receiver operating characteristic curve (AUC)} (closer to $50\%$ indicates weaker membership signal). For the generation task, we report both Top-1 accuracy and perplexity (PPL).

For the image-classification task, we use Top-1 accuracy as the primary utility metric.
We additionally evaluate \emph{black-box} membership leakage via a membership inference attack (MIA) adapted to unlearning~\cite{kurmanji2023towards}:
we form an attack dataset using the unlearned model's per-example output loss on forget-set samples (members) versus an identically distributed held-out test set (non-members), train an attack classifier on a class-balanced split, and report its performance on a disjoint held-out evaluation split using the \emph{area under the receiver operating characteristic curve (AUC)} (values closer to $50\%$ indicate a weaker membership signal).
This diagnostic measures distinguishability from \emph{model outputs} and is independent of the zero-knowledge privacy of the proof transcript.
For the generation task, we report both Top-1 accuracy and perplexity (PPL).

All main experiments in Section~\ref{sec:main_results} are conducted on a virtual machine with 32~vCPUs and 256~GB of RAM.  
To assess deployment feasibility on constrained hardware, Section~\ref{sec:edge_eval} reports additional results for ZK-SNARK proof generation executed on an \emph{iPhone~14~Pro~Max} (A17 chip, 6-core CPU).

\begin{table}[tb]
\centering
\caption{\textbf{Unlearning efficacy on LLM.}
Forgetting vs.\ personalization trade-off. Acc.\ is in \% (higher is better); PPL is perplexity (lower is better).}
\label{tab:llm_results}
\vspace{0.4em}
\resizebox{0.80\columnwidth}{!}{
\begin{tabular}{lcccc}
\toprule
\multirow{2}{*}{\textbf{Method}} & \multicolumn{2}{c}{\textbf{Forget}} & \multicolumn{2}{c}{\textbf{Personal}} \\
\cmidrule(lr){2-3}\cmidrule(lr){4-5}
& \textbf{Acc.} & \textbf{PPL} & \textbf{Acc.} & \textbf{PPL} \\
\midrule
Pre-unlearning & 65.3 & 6.89 & 77.4 & 3.16 \\
\midrule
Mask only (2\%) & 61.9 & \textbf{10.85} & 71.6 & 3.44 \\
\textbf{\name} & \textbf{61.8} & 10.62 & \textbf{75.6} & \textbf{3.20} \\
\midrule
\textit{Exact unlearning} & 61.5 & 10.53 & 77.3 & 3.28 \\
\bottomrule
\end{tabular}
}
\end{table}

\subsection{Main Results}
\label{sec:main_results}

We present the main empirical results addressing EQ1–EQ3, focusing on the forgetting–retention trade-off and the computational feasibility of ZKP generation.
The ablation study for EQ4, examining the sensitivity of key hyperparameters, is included in Appendix~\ref{sec:ablations}.

\textbf{Forgetting vs.\ personalization retention.}
Table~\ref{tab:main_results} reports the effect of different unlearning strategies on the personalized model’s performance, measured on held-out subsets of the \forgetset and personalization data.  
The pruning-based mask sharply reduces accuracy on the \forgetset while lowering personalized accuracy by about 3.4\%, highlighting the trade-off between forgetting and retention. 
\name recovers nearly 99\% of the personalized accuracy lost due to masking, while further suppressing the \forgetset accuracy.

% OLD: Applying the unlearning mask alone substantially reduces the model's accuracy on the \forgetclass (from 93.7\% to 60.5\%), indicating that the pruning successfully disrupts the model’s reliance on the targeted data. However, this step also introduces a non-negligible performance drop on the client’s personalized data (from 71.5\% to 69.4\%), underscoring the trade-off between forgetting and utility preservation. 
% Our weight adjustment method restores most of the lost personalized accuracy (up to 70.9\%) and further lowers the \forgetclass accuracy to 59.5\%. Measured relative to the masked model, these gains highlight our method’s ability to enhance forgetting while recovering utility. Notably, this is achieved by modifying only 2\% of the MLP parameters.

Compared to gradient-ascent and SCRUB baselines, both limited to a single epoch for ZK tractability, our approach achieves stronger forgetting, better retention, and MIA leakage.  
These results demonstrate that curvature-aware compensation effectively mitigates the loss of personalized utility caused by mask-based forgetting.

\textbf{ZK proof generation efficiency.}
Table~\ref{tab:main_results} summarizes the computational cost of generating and verifying ZKP for different unlearning operators.
In our deployment setting, the reported costs correspond to an \emph{unlearning window/event}: the provider batches deletion requests within a window, publishes a single public artifact $\Psi$ for that window, and each client generates one proof certifying $\theta_u = U(\theta_p;\Psi)$ for its locally personalized model (rather than producing a separate proof per individual deletion request).
For approximate (GA, SCRUB) and exact unlearning baselines, the gradient computation circuits exceeded available memory; following standard practice~\cite{zcash-halo2,maheri2025telesparse}, we partitioned the computation into multiple sub circuits, each capped at $2^{20}$~rows to fit within 256,GB RAM.
The per-sample proving time, proof size, and verification time were measured for each sub-circuit and then scaled by the number of sub-circuits per sample and by the total number of samples in each unlearning or learning procedure.
A brief note on batching and the resulting throughput/communication interpretation is provided in Appendix~\ref{sec:appendix_deployment_throughput}.

Our method achieves orders-of-magnitude lower proving time and memory usage than optimization-based baselines.
This improvement stems from its linear formulation, which avoids iterative updates and gradient reconstruction inside the circuit.
The results confirm that efficient, verifiable personalized unlearning is feasible for high-capacity models without compromising performance or proof succinctness.

\subsection{Edge-Device Evaluation}
\label{sec:edge_eval}

\begin{table}[tb]
\centering
\caption{\textbf{Edge-device proof overhead.}
Per-block (ViT-B/16) proving cost on the client device.}
\label{tab:edge_proof}

\setlength{\tabcolsep}{4pt}
\renewcommand{\arraystretch}{1.05}

\resizebox{0.95\columnwidth}{!}{%
\begin{tabular}{c|c|ccc}
\toprule
\textbf{Fisher block} & \textbf{$k$} & \textbf{Proving time} & \textbf{Peak mem.} & \textbf{Proof size} \\
\midrule
\multirow{3}{*}{\centering 256}
& 2\% & 0.61 & 87 & 34.6 \\
& 4\% & 1.12 & 96 & 34.9 \\
& 8\% & 1.58 & 98 & 35.6 \\
\midrule
\multirow{3}{*}{\centering 512}
& 2\% & 0.77 & 110 & 52.1 \\
& 4\% & 1.40 & 360 & 52.1 \\
& 8\% & 2.80 & 890 & 54.2 \\
\bottomrule
\end{tabular}%
}
\end{table}

To assess real-world feasibility, we measure proof-generation performance on an \emph{iPhone 14 Pro Max}.  
The prover runs locally on the device, while the verifier executes on a remote server.  
We record wall-clock proving time, peak memory usage, and proof size for a single block update under different pruning ratios in ~Table~\ref{tab:edge_proof}.

% \textbf{Results.}  
% Table~\ref{tab:edge_proof} summarizes the results.  
% Despite the computational constraints of mobile hardware, proof generation remains practical: end-to-end proving for a ViT-B/16 block completes within a few minutes, with memory well below device limits and proof sizes remaining compact.  
% These findings demonstrate that our zero-shot, ZK-friendly unlearning framework can be executed even on commodity mobile devices, indicating its applicability to realistic edge-deployment scenarios for relatively large models such as Vision Transformers.

% \paragraph{Discussion.}
% The results in Table~\ref{tab:edge_proof} show predictable scaling: proving time grows nearly linearly with the pruning ratio $k$ and sub-quadratically with the Fisher block size, while proof size remains almost constant. Although memory increases at higher sparsity, the overall footprint stays well within the device’s capacity. Because ZK-SNARK verification is lightweight, the model provider can efficiently validate proofs from many clients in parallel. Even for a relatively large model such as ViT-B/16, the proposed \name framework completes proof generation on a mobile processor within a few hours without requiring high memory, and can further reduce wall-clock time by parallelizing block proofs. These characteristics make the approach practical for edge intelligence deployments, such as on-device personalization in transformer-based Apple Intelligence~\cite{gunter2024apple}, where users can locally prove correct unlearning while the provider verifies many clients concurrently.

Despite the computational limits of mobile hardware, proof generation remains practical. As shown in Table~\ref{tab:edge_proof}, proving time scales nearly linearly with the pruning ratio $k$ and sub-quadratically with the Fisher block size, while memory and proof sizes stay within device capacity. For a ViT-B/16 block, end-to-end proof generation completes within a few hours without requiring high memory and can further reduce wall-clock time by parallelizing block proofs. Because ZK-SNARK verification is lightweight, the provider can validate proofs from many clients in parallel. These results confirm that the proposed \name framework is feasible for edge intelligence deployments, such as on-device personalization in transformer based Apple Intelligence~\cite{gunter2024apple}, where users can locally prove correct unlearning while the provider verifies multiple clients concurrently.

\section{Conclusion and Future Work}
\label{sec:conclusion}

This work introduced the first framework for \emph{verifiable personalized unlearning} on edge devices, motivated by the need to enforce user-data deletion under privacy and regulatory constraints without relying on trust in local computation. We proposed a pruning-based approximate unlearning algorithm with OBS compensation, designed to be \emph{zero-shot} (requiring no retraining iterations) and inherently compatible with ZKP systems. The approach enables providers to verify, in zero knowledge, that a client correctly executed an agreed unlearning transformation on its personalized model without revealing model weights or private data. Our \emph{linear operation formulation} avoids stochastic optimization inside the circuit, eliminating vulnerabilities linked to randomness in SGD and thereby improving robustness against forging attacks on verifiable unlearning.

Empirical results on personalized ViT models fine-tuned on ImageNet-Sketch confirm that the proposed method effectively removes \forgetset influence while recovering over $99\%$ of the personalized accuracy lost through naive pruning. These findings demonstrate that efficient, privacy-preserving verification of approximate unlearning is feasible even for large transformer architectures on mobile devices.

Several directions remain for future research. Our current formulation assumes that residual-gradient, cross-curvature, and quadratic effects on unmasked parameters are sufficiently small, so that compensation does not negate forgetting. One extension is to project compensation onto subspaces orthogonal to forget directions, canceling these terms while remaining computationally light and ZK-compatible. Exploring alternative proving systems, such as MPC-based or polynomial-commitment-based SNARKs, could further improve scalability and reduce proof latency. From a privacy perspective, incorporating differential privacy into mask construction may help defend against inversion and reconstruction attacks during verification~\cite{DBLP:conf/icml/ZhangCSL24,mahdiwarp}. At the systems level, hardware-assisted secure erasure would complement our cryptographic verification by ensuring that prior model states are permanently removed from device storage.

Another promising direction is to extend the framework beyond sample-defined forget sets to richer forms of forgetting. In particular, \emph{concept-level unlearning} aims to remove knowledge associated with semantic concepts rather than individual samples, while \emph{feature-level unlearning} targets finer-grained attributes or representation components. Adapting verifiable unlearning mechanisms to support these richer request types remains an open challenge. More broadly, it would be valuable to support additional deletion strategies within the same verifiable interface, provided they can be expressed as deterministic, circuit-checkable operators. Finally, when personalization and \forgetset distributions overlap substantially, balancing retention and forgetting becomes inherently difficult. Determining whether approximate unlearning suffices, or whether exact retraining remains necessary in such regimes, is an important open question for future work.

\nocite{langley00}

\bibliography{example_paper}
\bibliographystyle{mlsys2025}
\clearpage

\section*{ACKNOWLEDGEMENTS}
We wish to acknowledge the thorough and useful feedback from anonymous reviewers and our shepherd. The research in this paper was supported by the UKRI  Open Plus Fellowship (EP/W005271/1 Securing the Next Billion Consumer Devices on the Edge) and EU CHIST-ERA GNNs for Network Security and Privacy (GRAPHS4SEC) projects.
Alex Davidson's work is funded by national funds through FCT --- Fundação para a Ciência e a Tecnologia, I.P., under the project ``Fully-Homomorphic Encryption from Post-Quantum Code-Based Assumptions'', ref. 2024.12595.CMU, DOI \href{https://doi.org/10.54499/2024.12595.CMU}{10.54499/2024.12595.CMU}, and under the LASIGE Research Unit, ref. UID/00408/2025, DOI \href{https://dx.doi.org/10.54499/2024.07643.IACDC}{10.54499/2024.07643.IACDC}, and the LASIGE Research Unit, ref.\ UID/00408/2025 -- LASIGE.
% \end{acks}

%%%%%%%%%%%%%%%%%%%%%%%%%%%%%%%%%%%%%%%%%%%%%%%%%%%%%%%%%%%%%%%%%%%%%%%%%%%%%%%
%%%%%%%%%%%%%%%%%%%%%%%%%%%%%%%%%%%%%%%%%%%%%%%%%%%%%%%%%%%%%%%%%%%%%%%%%%%%%%%
% SUPPLEMENTAL CONTENT AS APPENDIX AFTER REFERENCES
%%%%%%%%%%%%%%%%%%%%%%%%%%%%%%%%%%%%%%%%%%%%%%%%%%%%%%%%%%%%%%%%%%%%%%%%%%%%%%%
%%%%%%%%%%%%%%%%%%%%%%%%%%%%%%%%%%%%%%%%%%%%%%%%%%%%%%%%%%%%%%%%%%%%%%%%%%%%%%%
\appendix
\section{Why the non-mask terms cannot undo unlearning}
\label{app:cross-comp-analysis}

\subsection{Goal and setup}

\paragraph{Objective.}
For the decomposition in \eqref{eq:forget-decomp}, prove that the \emph{non-mask} contribution on the complement block \(C\)---the sum of the residual linear term, the cross-curvature term, and the quadratic term on \(C\)---cannot be strongly negative; i.e., it cannot undo the increase produced by the mask-only part.

\paragraph{Anchor and notation.}
Let \(a:=\theta_p=\theta_0+BA\).
Define the \forgetset derivatives at \(a\):
\[
g:=\nabla_\theta L(a;D_f),\qquad H:=\nabla^2_\theta L(a;D_f).
\]
Block-partition by the mask support \(M\) and its complement \(C\):
\[
g=\begin{bmatrix}g_M\\ g_C\end{bmatrix},\qquad
H=\begin{bmatrix}H_{MM}&H_{MC}\\ H_{CM}&H_{CC}\end{bmatrix}.
\]
Masking enforces \((a+\delta W)_M=0\) via \(\delta W_m=(-a_M,0_C)\), where \(a_M:=(a)_M\).
The compensation \(\delta W_c\) is supported on \(C\) (i.e., \((\delta W_c)_M=0\)).

\paragraph{Quadratic model.}
Using \eqref{eq:taylor-second} at \(a\) and substituting \(\delta W=\delta W_m+\delta W_c\) yields
\begin{equation}
\Delta L_f \;\approx\; S_{\text{mask}} + f(\delta W_c) + R_3,
\label{eq:app-master}
\end{equation}
with
\begin{equation}
\begin{aligned}
S_{\text{mask}} &:= -\,g_M^\top a_M + \tfrac12\,a_M^\top H_{MM}a_M, \\
f(\delta W_c) &:= g_C^\top \delta W_c 
+ \delta W_c^\top H_{CM}(-a_M) \\
&\quad + \tfrac12\,\delta W_c^\top H_{CC}\delta W_c.
\end{aligned}
\label{eq:app-defs}
\end{equation}
Set
\begin{equation}
\begin{aligned}
b &:= g_C - H_{CM}a_M, \\
Q &:= H_{CC} + \lambda I \quad (\lambda>0\ \text{for damping, so } Q \succ 0).
\end{aligned}
\label{eq:app-bQ}
\end{equation}
so
\begin{equation}
f(\delta W_c) \;=\; b^\top \delta W_c \;+\; \tfrac12\,\delta W_c^\top Q\,\delta W_c.
\label{eq:app-f}
\end{equation}
The cubic remainder satisfies \(\|R_3\|\le c\,\|\delta W_m+\delta W_c\|^3\) and is small for sparse masks and damped compensation.

\subsection{Worst-case (most negative) analysis on \texorpdfstring{\(D_f\)}{Df}}

\paragraph{Lemma B.1 (completion of the square).}
For any \(Q\succ0\) and any \(x,y\),
\begin{equation}
x^\top y \;+\; \tfrac12\,y^\top Q\,y
\;=\;
\tfrac12\,\big\|Q^{1/2}(y+Q^{-1}x)\big\|_2^2 \;-\; \tfrac12\,\big\|Q^{-1/2}x\big\|_2^2.
\label{eq:app-cos}
\end{equation}
Applying \eqref{eq:app-cos} with \(x=b\), \(y=\delta W_c\) and using \eqref{eq:app-f} gives
\begin{equation}
f(\delta W_c)
\;=\;
\tfrac12\,\big\|Q^{1/2}(\delta W_c + Q^{-1}b)\big\|_2^2
\;-\;
\tfrac12\,\big\|Q^{-1/2}b\big\|_2^2.
\label{eq:app-f-cos}
\end{equation}

\paragraph{Corollary B.2 (most negative value).}
From \eqref{eq:app-f-cos},
\begin{equation}
\begin{aligned}
\min_{\delta W_c} f(\delta W_c)
&= -\,\tfrac12\,\|Q^{-1/2}b\|_2^2, \\
&\text{attained at}\quad \delta W_c^\star = -Q^{-1}b.
\end{aligned}
\label{eq:app-min}
\end{equation}

\paragraph{Bound B.3 (spectral control).}
Let \(\mu_{\min}(Q)\) be the smallest eigenvalue of \(Q\).
Then
\begin{equation}
\begin{aligned}
\|Q^{-1/2}b\|_2^2
&\le \frac{1}{\mu_{\min}(Q)}\,\|b\|_2^2 \\
&\le \frac{1}{\mu_{\min}(Q)}\Big(\|g_C\|_2 + \|H_{CM}\|\,\|a_M\|_2\Big)^2.
\end{aligned}
\label{eq:app-spectral}
\end{equation}
\emph{Derivation:} the first inequality is the Rayleigh bound; the second uses the triangle inequality and \(\|H_{CM}a_M\|_2\le \|H_{CM}\|\,\|a_M\|_2\).

\paragraph{Consequence.}
No choice of \(\delta W_c\) can reduce the mask-only increase by more than
\(\tfrac{1}{2\mu_{\min}(Q)}\big(\|g_C\|_2+\|H_{CM}\|\,\|a_M\|_2\big)^2\).
This bound tightens with stronger damping (larger \(\mu_{\min}(Q)\)), weaker cross-curvature \(\|H_{CM}\|\), smaller residual gradient \(\|g_C\|\) on \(C\), and moderate mask budgets (smaller \(\|a_M\|_2\)).

\subsection{Actual compensation used next (group-OBS on \texorpdfstring{\(D_p\)}{Dp})}

\paragraph{Setup.}
Let \(C_p\succ0\) be the OBS metric (e.g., damped empirical Fisher) and \(K:=C_p^{-1}\).
Group-OBS solves
\begin{equation}
\min_{\delta W}\ \tfrac12\,\delta W^\top C_p\,\delta W
\qquad
\text{s.t.}\quad
E_M^\top \delta W + a_M = 0,
\label{eq:app-obs-qp}
\end{equation}
with \(E_M=[e_i]_{i\in M}\).
The KKT solution is
\begin{equation}
\begin{aligned}
\delta W^{\mathrm{obs}} &\;=\; -\,K\,E_M\,(E_M^\top K E_M)^{-1} a_M, \\
\Rightarrow\quad
\delta W_c^{\mathrm{obs}} &\;=\; -A\,a_M, \quad
A := K_{C,M}\big(K_{M,M}\big)^{-1}.
\end{aligned}
\label{eq:app-obs-sol}
\end{equation}

\paragraph{Contribution on \(D_f\).}
Define
\begin{equation}
u:=Q^{-1/2}b,
\qquad
v:=Q^{1/2}A a_M.
\label{eq:app-uv}
\end{equation}
Substituting \(\delta W_c^{\mathrm{obs}}\) into \eqref{eq:app-f} and factoring by \(Q^{1/2}\) yields
\begin{equation}
f(\delta W_c^{\mathrm{obs}})
\;=\;
-\,u^\top v \;+\; \tfrac12\,\|v\|_2^2
\;=\;
\tfrac12\,\|v-u\|_2^2 \;-\; \tfrac12\,\|u\|_2^2.
\label{eq:app-obs-f}
\end{equation}
Hence,
\begin{equation}
\begin{aligned}
-\,\tfrac12\,\|u\|_2^2
&\;\le\;
f(\delta W_c^{\mathrm{obs}})
\;\le\;
\tfrac12\,\|v\|_2^2 + \|u\|_2\,\|v\|_2, \\
&\text{and}\quad
f(\delta W_c^{\mathrm{obs}})\ge 0
\ \text{whenever}\ \
\|v\|_2 \ge 2\|u\|_2.
\end{aligned}
\label{eq:app-obs-bounds}
\end{equation}
With
\begin{equation}
\begin{aligned}
\|u\|_2 &\le \frac{\|g_C\|_2 + \|H_{CM}\|\,\|a_M\|_2}{\sqrt{\mu_{\min}(Q)}},\\
\|v\|_2 &\le \|Q^{1/2}A\|\,\|a_M\|_2.
\end{aligned}
\label{eq:app-uv-bounds}
\end{equation}
damping in \(Q\) and \(C_p\) (which shrinks \(\|A\|\)), and moderate mask budgets make \(f(\delta W_c^{\mathrm{obs}})\) small (often nonnegative).

% \subsection{Sufficient condition for ``no undoing''}

% Combining \eqref{eq:app-master}--\eqref{eq:app-f} with the bounds above, a simple sufficient condition for guaranteed forgetting is
% \begin{equation}
% S_{\text{mask}} \;>\; \frac{1}{2\,\mu_{\min}(Q)}\Big(\|g_C\|_2+\|H_{CM}\|\,\|a_M\|_2\Big)^2 \;+\; |R_3|.
% \label{eq:app-sufficient}
% \end{equation}
% When the concrete compensation of the next subsection is used, the deviation from mask-only equals \(f(\delta W_c^{\mathrm{obs}})\) in \eqref{eq:app-obs-f} and remains bounded by \eqref{eq:app-obs-bounds}--\eqref{eq:app-uv-bounds}.

\paragraph{Takeaway.}
Combining \eqref{eq:app-master} with the bounds in \eqref{eq:app-f-cos}–\eqref{eq:app-uv-bounds} yields a robust lower bound
\(
\Delta L_f \ge S_{\text{mask}} - \tfrac{1}{2}\|Q^{-1/2}b\|_2^2 - |R_3|
\)
and, for the compensation used next, the explicit range in \eqref{eq:app-obs-bounds}.
With standard damping and modest residual gradient/cross-curvature (i.e., small \(\|g_C\|\) and \(\|H_{CM}\|\)) and moderate mask budgets, the non-mask terms remain uniformly controlled and do not materially offset the mask-only increase.

\subsection{How each equation follows}

\begin{itemize}
\item \eqref{eq:forget-decomp} is the blockwise expansion of \eqref{eq:taylor-second} with \(\delta W=\delta W_m+\delta W_c\).
\item \eqref{eq:app-defs}–\eqref{eq:app-f} define the mask-only and non-mask terms, isolate \(b\), and fold damping into \(Q\).
\item Lemma~\ref{eq:app-cos} is the standard completion-of-square identity for SPD \(Q\).
\item \eqref{eq:app-min} follows by minimizing \eqref{eq:app-f-cos}.
\item \eqref{eq:app-spectral} uses the Rayleigh quotient and submultiplicativity.
\item \eqref{eq:app-obs-qp}–\eqref{eq:app-obs-sol} are the KKT solution of a strictly convex quadratic program; the \(C\)-block form is the Schur complement.
\item \eqref{eq:app-obs-f} comes from substituting \(-A a_M\) into \(f(\cdot)\) and factoring by \(Q^{1/2}\).
\item The norm bounds in \eqref{eq:app-obs-bounds}–\eqref{eq:app-uv-bounds} follow from operator norms and \(\mu_{\min}(Q)\).
\end{itemize}

\section{Background: Circuits and Zero-Knowledge Proof Systems}
\label{sec:zk_background}

Modern ZKP systems enable one party (the prover) to convince another (the verifier) that a computation was executed correctly without revealing private inputs or intermediate states. 
For instance, in the context of verifiable unlearning, the computation corresponds to the application of the unlearning operator \(U(\cdot)\) on a personalized model. 
Below we outline the key building blocks of ZK-SNARKs, from circuit representation to proof generation.

\point{Arithmetic circuits}
Any deterministic computation can be expressed as an \emph{arithmetic circuit} over a finite field~\(\mathbb{F}\). 
The circuit is a directed acyclic graph composed of addition and multiplication gates whose wires carry field elements.
Each wire represents an intermediate variable, and a satisfying \emph{witness} is an assignment of values to these wires that makes all gate constraints valid.

\point{From circuits to R1CS}
To check circuit correctness algebraically, each multiplication gate is converted into a \emph{rank-1 constraint} of the form 
\(\langle \mathbf{a}, \mathbf{s} \rangle \cdot \langle \mathbf{b}, \mathbf{s} \rangle = \langle \mathbf{c}, \mathbf{s} \rangle\),
where \(\mathbf{s}\) denotes all wire values. 
A set of \(m\) gates yields \(m\) such constraints, collectively called a Rank-1 Constraint System (R1CS). 
Satisfying all R1CS equations is equivalent to evaluating the original circuit correctly.

\point{From R1CS to QAP}
The R1CS constraints are further encoded into a single \emph{Quadratic Arithmetic Program} (QAP), which represents all constraints as a polynomial identity. 
Let \(t(x)\) be a target polynomial vanishing at predetermined points; the prover constructs polynomials \(U(x), V(x), W(x)\) from the circuit such that
\[
U(x)V(x) - W(x) = H(x)t(x).
\]
This holds if and only if the witness satisfies the entire circuit, thereby transforming constraint satisfaction into a single polynomial divisibility condition~\cite{groth2016size}.

\point{From QAP to succinct proofs}
zk-SNARK constructions such as Pinocchio and Groth16 use homomorphic commitments and bilinear pairings to let the verifier check the above polynomial identity at a random evaluation point without learning any private information.
This reduction allows the verifier to confirm that a prover knows a valid witness with only a few cryptographic checks, yielding \emph{succinct} proofs and constant verification time. 
The circuit description is fixed and public; the witness, encoding private model parameters or data, remains hidden.

\point{Halo2 and modern polynomial commitments}
Halo2~\cite{zcash-halo2} extends the above paradigm using transparent polynomial commitment schemes (e.g., KZG-type) that avoid trusted setup and support recursive proofs. 
It represents circuits as constraint systems over field polynomials and verifies each constraint through low-degree testing. 
Halo2 thus achieves scalability while preserving soundness, correctness, and zero-knowledge under standard cryptographic assumptions. 
These three properties, informally ensure that proofs are both reliable and privacy-preserving:
\begin{itemize}[leftmargin=*]
    \item \textbf{Soundness:} a malicious prover cannot convince the verifier of a false statement;
    \item \textbf{Correctness:} an honest prover can always convince the verifier of a true statement;
    \item \textbf{Zero-knowledge:} the verifier learns nothing beyond the validity of the statement.
\end{itemize}
 
Formal definitions of these properties, together with the binding and hiding guarantees of the underlying commitment scheme, are provided in Appendix~\ref{sec:appendix_all_properties}.

\point{Relevance to our framework}
In our setting, the circuit encodes the unlearning transformation \(U(\cdot)\), the public inputs correspond to the provider’s traceability artifacts \(\Psi\), and the private witness includes the client’s model parameters $\theta_p$ and personalization data. 
Halo2’s modular arithmetic-circuit abstraction allows efficient verification of sparse linear updates such as our Group-OBS compensation, making it a suitable foundation for verifiable personalized unlearning.

\section{Formal Properties of Halo2 and Commitment Schemes}
\label{sec:appendix_all_properties}

Let \(\kappa\) denote the security parameter, and let \(\nu(\kappa)\) be a negligible function.  
Below we restate the formal guarantees provided by the Halo2 proving system, together with the standard properties of its underlying polynomial commitment scheme.

\colonpoint{Soundness.}
A proving system is \textbf{sound} if no efficient (possibly malicious) prover can convince the verifier of a false claim, except with negligible probability.  
Formally, for every probabilistic polynomial-time (PPT) prover \( \mathcal{P}^* \) and any statement \( \phi \notin L \),
\begin{equation}
\label{eq:zk_soundness}
\Pr[\langle \mathcal{P}^*, \mathcal{V} \rangle (\phi) = 1 \mid \phi \notin L] \leq \nu(\kappa).
\end{equation}

\colonpoint{Correctness.}
A proving system is \textbf{correct} if an honest prover can always convince the verifier of a true statement.  
For any valid instance \( \phi \in L \) and witness \( w \),
\begin{equation}
\Pr[\langle \mathcal{P}, \mathcal{V} \rangle (\phi, w) = 1] = 1 - \nu(\kappa).
\end{equation}

\colonpoint{Zero-Knowledge.}
A proving system satisfies the \textbf{zero-knowledge} property if the verifier learns nothing beyond the validity of the proven statement.  
Formally, for any PPT verifier \( \mathcal{V}^* \), there exists a simulator \( \mathcal{S} \) that can produce an indistinguishable view:
\begin{equation}
\text{View}(\mathcal{P}(w), \mathcal{V}^*(\phi)) \approx \mathcal{S}(\phi).
\end{equation}

\medskip
\noindent
Halo2 implements a polynomial commitment scheme (e.g., KZG-type) that enables succinct and verifiable polynomial evaluations, while maintaining confidentiality of witness data.  
Such schemes satisfy two essential security properties: \emph{binding} and \emph{hiding}.

\colonpoint{Binding.}
The \textbf{binding} property ensures that a prover cannot open a single commitment to two distinct values.  
Let \( \operatorname{Comm}:\mathbb{F}^n\rightarrow\mathcal{C} \) denote the commitment function.
For any two distinct vectors \( \mathbf{v}, \mathbf{v'} \in \mathbb{F}^n \), it is computationally infeasible to find randomizers \( r,r' \) such that both open to the same commitment:
\begin{equation}
\begin{aligned}
\forall\, \text{PPT algos } \mathcal{A},\quad
&\Pr\!\Big[
(c,\mathbf{v},\mathbf{v'}) \leftarrow \mathcal{A}() \;\Big|\;
\operatorname{Comm}(\mathbf{v};r)=c,\\
&\operatorname{Comm}(\mathbf{v'};r')=c,\;
\mathbf{v}\neq\mathbf{v'}
\Big]
\;\leq\;
\nu(\kappa).
\end{aligned}
\label{eq:binding_property}
\end{equation}
\colonpoint{Hiding.}
A commitment scheme is \textbf{hiding} if the committed value remains computationally indistinguishable to any adversary without the opening randomness.  
For any \( \mathbf{v}, \mathbf{v'} \in \mathbb{F}^n \) and independent random coins \( r, r' \), the distributions of \( \operatorname{Comm}(\mathbf{v}; r) \) and \( \operatorname{Comm}(\mathbf{v'}; r') \) are indistinguishable to all PPT adversaries \( \mathcal{A} \):
\begin{equation}
\label{eq:commitment-hiding}
\begin{aligned}
\Big|
\Pr[\mathcal{A}(\operatorname{Comm}(\mathbf{v};r))&=1] -\;
\\[-3pt]
\Pr[\mathcal{A}(\operatorname{Comm}(\mathbf{v'}&;r'))=1]
\Big|
\;\leq\;
\nu(\kappa).
\end{aligned}
\end{equation}

\noindent
Together, these properties—soundness, correctness, zero-knowledge, binding, and hiding—guarantee that proofs in our framework (built upon Halo2) are both verifiable and privacy-preserving under standard cryptographic assumptions.

\section{Dimensionality and Computational Cost Analysis}
\label{sec:appendix_dim_cost}

Let \(d\) denote the total number of model parameters, \(k = |M|\) the number of masked parameters, and \(\{d_b\}\) the dimensions of the Fisher blocks.  
Each block \(C_p^{(b)} \in \mathbb{R}^{d_b \times d_b}\) operates on its corresponding parameter slice \(\delta w^{(b)} \in \mathbb{R}^{d_b}\), with Lagrange multipliers \(\lambda_M \in \mathbb{R}^k\).  

Verifying the KKT stationary condition (\ref{eq:zk-stat}) requires one matrix–vector multiplication per block.  
This incurs a cost of \(\mathcal{O}(d_b^2)\) for dense blocks or \(\mathcal{O}(\mathrm{nnz}(C_p^{(b)}))\) when the curvature is sparse or structured.  
The mask-feasibility (Equation~\ref{eq:zk-feas}) and assembly (Equation~\ref{eq:zk-assembly}) constraints add a smaller \(\mathcal{O}(k + d)\) overhead.  

Hence, the dominant computational effort arises from the blockwise products \(C_p^{(b)} \delta w^{(b)}\), giving an overall asymptotic complexity of
\[
\sum_b \mathcal{O}(d_b^2) \approx \mathcal{O}(\mathrm{nnz}(C_p) + d + k).
\]
Because all constraints are linear, the resulting circuit is highly efficient, requiring only lightweight arithmetic operations and supporting succinct proof generation even on memory-constrained edge devices.

\begin{algorithm}[t]
\small
\caption{Verifiable Approximate Unlearning (Group-OBS, block-wise Fisher)}
\label{alg:verifiable_unlearning}
\begin{algorithmic}
\STATE \textbf{Input:} Public traceability artifact $\Psi=(m^\star,M)$; public commitments $\mathsf{Com}(\theta_p)$, $\mathsf{Com}(C_p)$
\STATE \textbf{Output:} Public commitment $\mathsf{Com}(\theta_u)$; ZK-SNARK proof $\pi$
\vspace{2pt}

\STATE \textbf{Client (offline, once):} Compute block-wise empirical Fisher $C_p$ on a small subsample of $D_p$ (first-order gradients only); publish $\mathsf{Com}(C_p)$.
\vspace{2pt}

\STATE \textbf{Unlearning request:} Provider sends $\Psi=(m^\star,M)$.
\vspace{2pt}

\STATE \textbf{Local unlearning (client):}
\begin{enumerate}
\item Form the selector $E_M=[e_i]_{i\in M}$ and extract $w_{p,M}=E_M^\top \theta_p$ (privately).
\item Compute the Group-OBS update (closed form):
\[
\delta w = -\,C_p^{-1}\,E_M\,\big(E_M^\top C_p^{-1}E_M\big)^{-1}\,w_{p,M}.
\]
\item Assemble the unlearned model: $\theta_u \leftarrow \theta_p + \delta w$.
\end{enumerate}
\vspace{2pt}

\STATE \textbf{Proof generation (client):} Produce a ZK-SNARK $\pi$ attesting that private witnesses
$\theta_p,\theta_u,\delta w,\lambda_M,C_p$ satisfy the linear KKT certificates:
\begin{align*}
&\text{(Assembly)}\qquad \qquad \theta_u = \theta_p + \delta w,\\
&\text{(Mask feasibility)}\quad E_M^\top \delta w + w_{p,M} = 0,\;\; w_{p,M}=E_M^\top \theta_p,\\
&\text{(KKT stationarity)}\quad C_p\,\delta w + E_M\,\lambda_M = 0,
\end{align*}
and that these openings match $\mathsf{Com}(\theta_p)$, $\mathsf{Com}(C_p)$, and define $\mathsf{Com}(\theta_u)$.
\vspace{2pt}

\STATE \textbf{Verification (provider):} Check $\pi$ against public inputs $(\Psi,\mathsf{Com}(\theta_p),\mathsf{Com}(C_p),\mathsf{Com}(\theta_u))$; accept iff valid.
\end{algorithmic}
\end{algorithm}

\section{Algorithm Pseudocode}
\label{sec:appendix_algorithm}

Algorithm~\ref{alg:verifiable_unlearning} outlines the complete procedure for our verifiable personalized unlearning framework, summarizing the client-side unlearning, compensation, and zero-knowledge proof generation steps described in Sections~\ref{section:unlearning_method} and~\ref{section:proof_generation_method}.

\section{Notation Table}
\label{sec:appendix_notation_table}
For a list of mathematical notations, please refer to Table~\ref{tab:notation_table}.

%%%%%%%%%%%%%%%%%%%%%%%%%%%%%%%%%%%%%%%%%%%%%%%%%%%%%%%%%%%%%%%%%%%%%%%%%%%%%%%
%%%%%%%%%%%%%%%%%%%%%%%%%%%%%%%%%%%%%%%%%%%%%%%%%%%%%%%%%%%%%%%%%%%%%%%%%%%%%%%

\section{Ablations and Sensitivity Study}
\label{sec:ablations}

% We ablate secondary terms in the unlearning formulation and study sensitivity to key hyperparameters.  
% Results are averaged over five runs using the same ViT-B/16 personalized setup as in Section~\ref{sec:main_results}, with a default 2\%.

To address EQ4, we ablate study sensitivity to key hyperparameters.  
Results are averaged over the top-5 performing runs using the same ViT personalized setup as in Section~\ref{sec:main_results}.

\begin{figure}[t]
    \centering
    \includegraphics[width=0.99\linewidth]{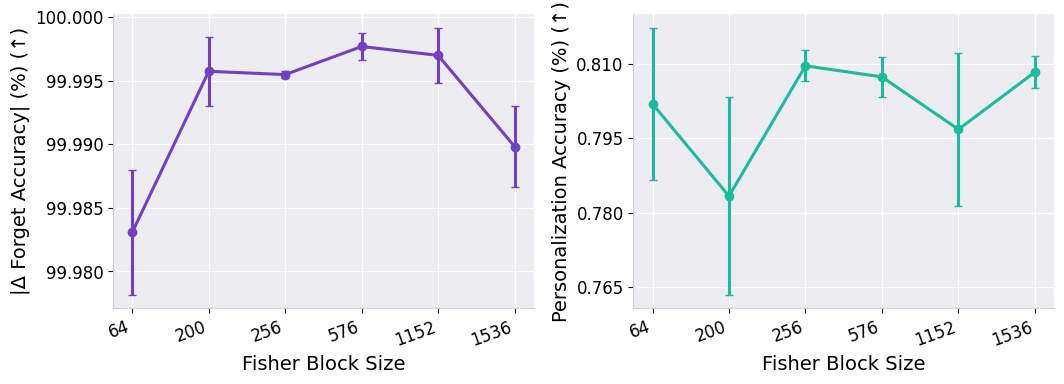}
    \caption{
        \textbf{Effect of Fisher block size.}
        Larger block sizes improve curvature stability during unlearning.
    }
    \label{fig:block_sensitivity}
\end{figure}

\begin{figure}[t]
    \centering
    \includegraphics[width=0.99\linewidth]{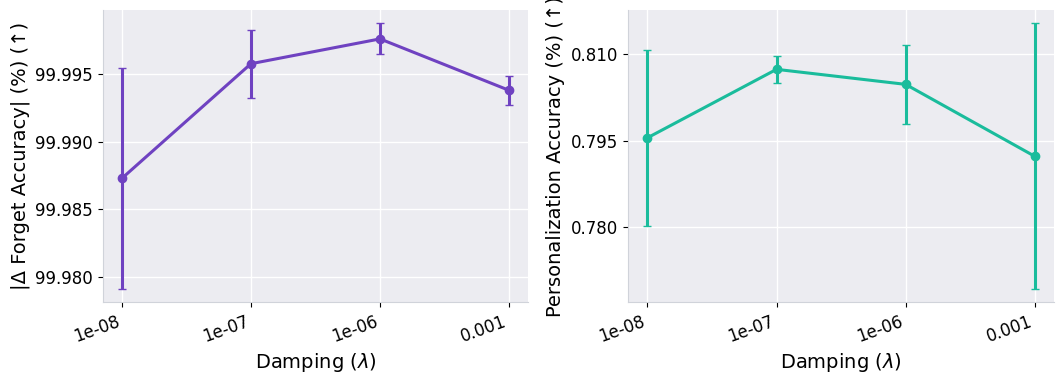}
    \caption{
        \textbf{Effect of damping coefficient.}
        Moderate damping yields the best trade-off between numerical stability
        and precision in curvature compensation.
    }
    \label{fig:damping_sensitivity}
\end{figure}

\begin{figure}[t]
    \centering
    \includegraphics[width=0.99\linewidth]{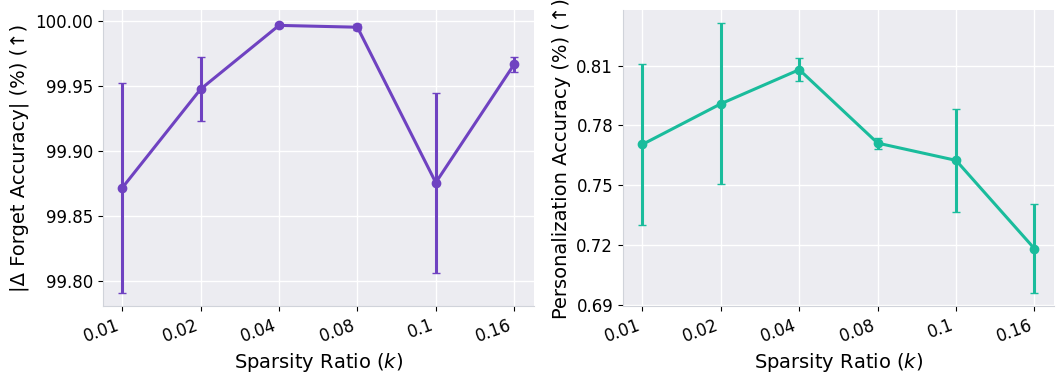}
    \caption{
        \textbf{Effect of sparsity ratio.}
        Higher sparsity enhances proof efficiency while maintaining strong unlearning performance.
    }
    \label{fig:sparsity_ratio_ablation}
\end{figure}

\section{Deployment model and throughput/communication interpretation}
\label{sec:appendix_deployment_throughput}

We consider a windowed deployment in which deletion requests are accumulated over a fixed interval and aggregated into a \forgetset $D_f$.
For each window, the provider publishes a single public traceability artifact $\Psi$ (e.g., a sparse mask), and each client produces a proof certifying that its released update satisfies $\theta_u = U(\theta_p;\Psi)$ for its locally personalized model.
Accordingly, the costs in Table~\ref{tab:main_results} should be interpreted at the granularity of one client proof per window.

A minimal systems interpretation follows from two measured quantities in Table~\ref{tab:main_results}: proof size $S$ and verification time $t_{\mathrm{ver}}$.
On the client side, the upload time under uplink bandwidth $b$ (bits/s) is
\[
t_{\mathrm{up}} \approx \frac{8S}{b}.
\]
For the ViT-B/16 configuration in Table~\ref{tab:main_results} ($S\approx 401$\,MB), this corresponds to approximately $2$--$3$ minutes at $b=20$--$25$\,Mbps.
On the verifier side, a single verifier instance has throughput
\[
\lambda_{\mathrm{ver}} \approx \frac{1}{t_{\mathrm{ver}}},
\]
so with $t_{\mathrm{ver}}\approx 10$ minutes the instance processes roughly one proof per $\sim 10$ minutes; throughput scales linearly with the number of verifier instances.

The reported $S$ and $t_{\mathrm{ver}}$ are totals across all verified model blocks.
Since our prover is block-decomposed, block proofs can be generated in parallel and streamed as they are produced, allowing overlap between proving and transmission and improving wall-clock latency in practice.
If communication becomes the dominant bottleneck, recursive composition can aggregate block proofs into a single succinct proof to reduce transmitted proof material; we view this as a systems optimization direction and report unaggregated costs in Table~\ref{tab:main_results}.

\section{Beyond Full Fine-Tuning: AdaptFormer Personalization}
\label{sec:appendix-adaptformer}

In the main ViT experiments (Section~\ref{sec:main_results}), we instantiate client-side personalization using full fine-tuning. However, \name is defined as a feasible, deterministic unlearning operator $U(\theta_p;\Psi)$ acting on the \emph{resulting} personalized parameters (Section~\ref{sec:problem_formulation}): the provider publishes a public traceability artifact $\Psi$ (a sparse mask), and the client computes a curvature-aware compensation from $D_p$ (Group-OBS) and proves operator compliance in zero knowledge. Since this workflow depends on the personalized model $\theta_p$ but not on the specific optimization procedure used to obtain $\theta_p$ from $\theta_0$, it should extend to other widely used personalization methods. To validate this claim, we repeat the ViT pipeline while changing only the personalization operator $P$: instead of full fine-tuning, we personalize ViT-B/16 on ImageNet-Sketch using AdaptFormer~\cite{chen2022adaptformer}, while keeping the unlearning configuration identical to Section~\ref{sec:exp_setup} (mask computed on the pretrained model, targeting MLP sublayers across Transformer blocks, with the same pruning ratio and the same Group-OBS compensation and ZK statement).

AdaptFormer~\cite{chen2022adaptformer} is a parameter-efficient adaptation method for vision transformers that freezes the pretrained backbone and injects lightweight bottleneck ``adapter'' modules into Transformer blocks (typically as an additional residual branch alongside the block MLP/FFN), training only the adapter parameters. A typical adapter computes a down-projection to a low-dimensional bottleneck, applies a nonlinearity, and projects back to the model dimension, which is then added (with a learnable scale) to the block representation, enabling scalable adaptation with a small number of trainable parameters relative to full fine-tuning. Table~\ref{tab:adaptformer-personalization} reports the forgetting--personalization trade-off under AdaptFormer personalization, using the same metrics as Table~1 but focusing on accuracy: ``Pre-unlearning model'' refers to the AdaptFormer-personalized $\theta_p$ prior to unlearning, ``Exact unlearning'' is the gold-standard retrain-on-retain then re-personalize using the same AdaptFormer operator $P$ (Equation~\ref{eq:counterfactual}), and \name applies our masked unlearning plus Group-OBS compensation starting from the AdaptFormer-personalized model.

\begin{table}[!t]
\centering
\small
\caption{\textbf{Unlearning efficacy on ViT under AdaptFormer personalization.}
Comparison of the forgetting--personalization trade-off.}
\label{tab:vit_adaptformer_results}
\vspace{0.4em}
\resizebox{\columnwidth}{!}{
\begin{tabular}{lcc}
\toprule
\multirow{2}{*}{\textbf{Method}} & \multicolumn{2}{c}{\textbf{Unlearning Efficacy}} \\
\cmidrule(lr){2-3}
& \textbf{Forget Acc. (\%)} $\downarrow$ & \textbf{Personal Acc. (\%)} $\uparrow$ \\
\midrule
Pre-unlearning & 92.5 & 82.5 \\
\midrule
Mask only (4\%) & 54.8 & 77.9 \\
\textbf{\name} & \textbf{54.7} & \textbf{81.4} \\
\midrule
\textit{Exact unlearning} & 54.7 & 82.3 \\
\bottomrule
\end{tabular}
\label{tab:adaptformer-personalization}
}
\end{table}

Under AdaptFormer personalization (Table~\ref{tab:adaptformer-personalization}), we observe the same qualitative behavior as with full fine-tuning.
Mask-only unlearning attains comparable forgetting but incurs an approximately \textbf{4.4-point} drop in personalization accuracy relative to the \textit{exact unlearning} gold standard.
In contrast, \name recovers about \textbf{3.5 points} of this gap (roughly \textbf{80\%} recovery), while matching the gold standard on forgetting.
Overall, this is consistent with the full fine-tuning results: \name substantially narrows the retention gap between naive masking and retrain-and-repersonalize, supporting its applicability across different personalization algorithms.

\section{Structured Forget Sets: Per-Class and Few-Class Unlearning}
\label{sec:appendix-classUnl}

In the main evaluation, we construct the \forgetset $D_f$ by sampling examples uniformly at random from the designated source split (Section~6.3), which typically yields a \emph{class-diverse} \forgetset. Real unlearning requests can be more \emph{structured} and thus more correlated, e.g., when a request targets data associated with a particular class or a small set of related labels. To probe this regime, we consider \emph{class-conditional} forget sets where $D_f$ is restricted to fewer classes, increasing intra-\forgetset correlation. Concretely, we evaluate two settings on ViT: (i) \textbf{1-class} forgetting, where $D_f$ consists of 600 examples from a single class, and (ii) \textbf{8-class} forgetting, where $D_f$ consists of 600 examples from each of 8 distinct classes (4{,}800 total). To enable comparisons across different \forgetset constructions, Table~\ref{tab:structured_forget_rel} reports the \emph{relative} deviation of \name from the \textit{Exact unlearning} gold standard, computed as $(\mathrm{metric}(\name)-\mathrm{metric}(\textit{Exact}))/\mathrm{metric}(\textit{Exact})$ for both forget accuracy and personalization accuracy; the \textit{Random} row corresponds to the main setting (Table~\ref{tab:main_results}). With this normalization, negative values for forget accuracy indicate \emph{stronger} forgetting than the gold standard, while positive values for personalization accuracy indicate \emph{better} retention.

For these structured forget sets, we keep the provider-side masking procedure fixed (i.e., the same masking construction and selection rule as in the main experiments) and only retune the \emph{client-side compensation} hyperparameters. This aligns with the provider--client workflow in our threat model: the provider publishes a single public traceability artifact $\Psi$ that is not customized per client, whereas each client can locally choose compensation settings to best preserve its personalization utility under the imposed unlearning request. This protocol isolates the impact of increased \forgetset structure on the forgetting--retention trade-off without relying on per-client tuning of the masking stage.

\begin{table}[!t]
\centering
\small
\caption{\textbf{Structured forget sets: relative deviation from exact unlearning.}
Relative differences (\%) between \name and \textit{Exact unlearning} across \forgetset constructions.}
\label{tab:structured_forget_rel}
\vspace{0.4em}
\resizebox{0.9\columnwidth}{!}{
\begin{tabular}{lcc}
\toprule
\multirow{2}{*}{\textbf{Forget-set}} & \multicolumn{2}{c}{\textbf{Relative diff.\ to Exact (\%)}} \\
\cmidrule(lr){2-3}
& $\boldsymbol{\Delta_{\mathrm{forget}}}$ ($\downarrow$) & $\boldsymbol{\Delta_{\mathrm{personal}}}$ ($\uparrow$) \\
\midrule
Random (Table~\ref{tab:main_results}) & +~0.20 & -~0.12 \\
8-class & +~0.25 & +~0.12 \\
1-class & 0.00 & +~0.37 \\
\bottomrule
\end{tabular}
}
\end{table}

As shown in Table~\ref{tab:structured_forget_rel}, \name remains tightly aligned with the \textit{Exact unlearning} gold-standard under structured (class-conditional) forget sets, and the alignment is slightly stronger when the forget set is concentrated in fewer classes. A plausible explanation is that increasing intra-$D_f$ correlation makes the forget signal more spatially and semantically localized in the representation space, so the provider mask can more cleanly isolate parameters tied to the targeted content while leaving client-relevant directions intact. Interestingly, in the 8-class and especially 1-class regimes, the masked update with compensation not only preserves personalization relative to the gold-standard but slightly improves it, consistent with the broader observation that mild sparsification/pruning can act as an implicit regularizer and occasionally improve generalization (and thus accuracy) rather than only compressing the model~\cite{bartoldson2020generalization,frankle2018lottery}.

% \section{Sensitivity to forget-set Size}
% In a similar vein, to measure the sensitivity to \forgetset sizes, larger forget sets were used. 600 examples from 108, 216, and 433 classes were chosen. t
% \label{sec:appendix-bigforget}

\section{Sensitivity to Forget-Set Size}
\label{sec:appendix-forgetsize}

In the main ViT evaluation (Table~\ref{tab:main_results}), we consider a default \forgetset size of $|D_f|/|D|=2.6\%$. In practice, unlearning requests may be substantially larger, and it is important to understand how the forgetting--personalization trade-off scales as a larger fraction of the training data is removed. To this end, we increase the \forgetset fraction from the default setting up to $20\%$ by sampling $D_f$ uniformly at random from the same source split as in Section~6.3, and evaluate \name across these regimes. For each \forgetset fraction, we compare against the \textit{Exact unlearning} oracle (retrain on $D\setminus D_f$ and then re-personalize) and report results using the same metrics as the main ViT experiments.

\begin{figure}[t]
  \centering
  \includegraphics[width=\columnwidth]{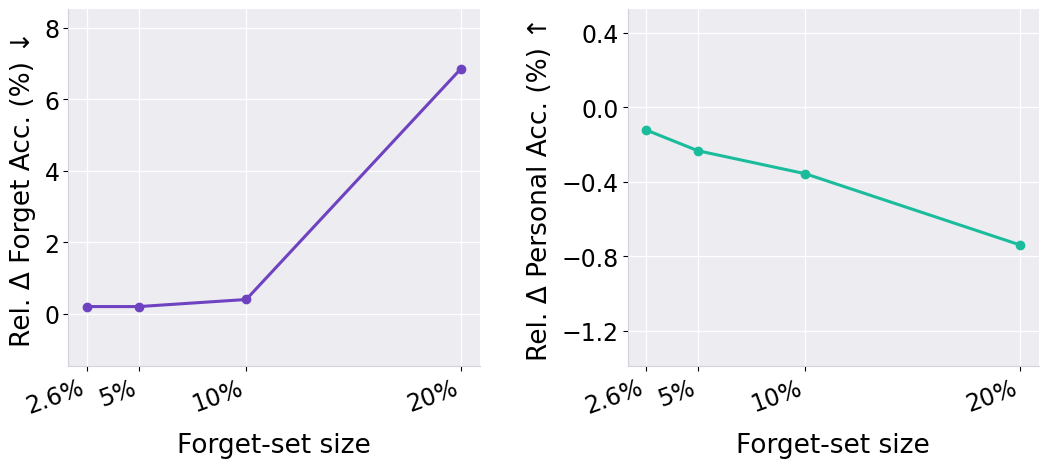}
  \vspace{-0.5em}
    \caption{\textbf{Scaling $|D_f|$: relative forget accuracy vs.\ exact unlearning.}
    Each point corresponds to a different \forgetset fraction (up to $20\%$). Lower (more negative) values indicate stronger forgetting relative to the \textit{Exact unlearning} oracle.}
  \label{fig:forgetsize_forget}
  \vspace{-0.8em}
\end{figure}

To make comparisons across different $|D_f|$ values scale-free, we plot the \emph{relative deviation} of \name from \textit{Exact unlearning} for both forget accuracy and personalization accuracy, using the same normalization as in Section~\ref{sec:appendix-classUnl}. Concretely, for a metric $m(\cdot)$ we compute
\[
\Delta_m(\%) \;=\; 100 \cdot \frac{m(\name)-m(\textit{Exact})}{m(\textit{Exact})}.
\]
We report $\Delta_{\mathrm{forget}}$ for forget accuracy and $\Delta_{\mathrm{personal}}$ for personalization accuracy. Under this convention, negative $\Delta_{\mathrm{forget}}$ indicates stronger forgetting than the oracle (lower forget accuracy), while positive $\Delta_{\mathrm{personal}}$ indicates better retention than the oracle (higher personalization accuracy). Figure~\ref{fig:forgetsize_forget} summarizes how these relative deviations evolve as the \forgetset fraction increases up to $20\%$.

Across increasing \forgetset sizes (up to $20\%$ of training data), personalization remains relatively stable: the relative deviation in personalization accuracy between \name and \textit{Exact unlearning} stays below $1\%$ in all cases, indicating that the client-side compensation continues to preserve personalized utility as $|D_f|$ grows. On the forgetting side, \name closely tracks the oracle for small-to-moderate forget ratios (up to $10\%$), with only negligible relative differences. The only clear degradation appears at the largest forget ratio ($20\%$), where forgetting becomes noticeably weaker. This behavior is consistent with our protocol in which the provider-side masking hyperparameter is fixed (e.g., $k=4\%$): for very large unlearning requests, a fixed mask budget can become a bottleneck, limiting how much of the oracle’s unlearning update can be expressed, whereas for smaller requests it remains sufficient to match the oracle while maintaining retention.

% \begin{figure}[t]
%   \centering
%   \includegraphics[width=\columnwidth]{Images/damping_ablation.png}
%   \vspace{-0.5em}
%   \caption{\textbf{Scaling $|D_f|$: relative personalization accuracy vs.\ exact unlearning.}
%   Each point corresponds to a different \forgetset fraction (up to $20\%$). Values report
%   $\Delta_{\mathrm{personal}}(\%) = 100\cdot(\mathrm{PersonalAcc}(\name)-\mathrm{PersonalAcc}(\textit{Exact}))/\mathrm{PersonalAcc}(\textit{Exact})$;
%   higher indicates better retention relative to the oracle.}
%   \label{fig:forgetsize_personal}
%   \vspace{-0.8em}
% \end{figure}

\begin{table*}[t]
\small
\centering
\caption{Notation. Vectors/matrices are real-valued; $d$ is the number of parameters.}
\label{tab:notation_table}
{\renewcommand{\arraystretch}{1.25}% ← 25% more vertical space
\setlength{\tabcolsep}{6pt}% (optional) tweak horizontal padding
}
\begin{tabular}{llp{0.65\columnwidth}}
\toprule
\multicolumn{3}{l}{\textbf{Data, sets, and distributions}}\\
$\mathcal{Z}=\mathcal{X}\times\mathcal{Y}$ & set & input–label space \\
$D=\{z_i\}_{i=1}^N$ & set & pretraining corpus \\
$D_f,\ D_r$ & set & \forgetset; retain set ($D_r=D\setminus D_f$) \\
$D_p$ & set & client’s personalization set \\
$p(\cdot\mid x;\theta)$ & dist. & predictive distribution of model $\theta$ \\
$d(\cdot,\cdot)$ & func. & divergence between predictive dists. (KL\textsubscript{fwd} in experiments) \\
$\mathcal{A}_p,\ \mathcal{A}_f$ & scalar & alignment on $D_p$, $D_f$ (Eq.~\eqref{eq:alignments}) \\
\midrule
\multicolumn{3}{l}{\textbf{Models and parameters}}\\
$\theta\in\mathbb{R}^d$ & vec. & model parameters \\
$\theta_0$ & vec. & pretrained/global model (ERM on $D$) \\
$\theta_p=P(\theta_0;D_p)$ & vec. & personalized model (via operator $P$) \\
$\theta_r^\star$ & vec. & retrained model on $D_r$ \\
$\theta^\star=P(\theta_r^\star;D_p)$ & vec. & gold-standard (exact) personalized model \\
$\theta_u$ & vec. & unlearned personalized model (output) \\
$\Delta_p=BA$ & mat./ten. & low-rank personalization update (rank $\le r$; conceptual) \\
$r$ & int & rank budget for $\Delta_p$ (when used) \\
\midrule
\multicolumn{3}{l}{\textbf{Losses and curvature}}\\
$\ell(\theta;z)$ & func. & per-example loss \\
$L(\theta;D)=\frac{1}{|D|}\sum_{z\in D}\ell(\theta;z)$ & func. & empirical risk on $D$ \\
$g_f(\theta)=\nabla_\theta L(\theta;D_f)$ & vec. & gradient on \forgetset \\
$g_p(\theta)=\nabla_\theta L(\theta;D_p)$ & vec. & gradient on personalization set \\
$H_f(\theta)=\nabla^2_\theta L(\theta;D_f)$ & mat. & Hessian on \forgetset \\
$H_p(\theta)=\nabla^2_\theta L(\theta;D_p)$ & mat. & Hessian on personalization set \\
$F_p(\theta)$ & mat. & empirical Fisher on $D_p$ \\
$C_p=F_p(\theta_p)+\lambda I\succ0$ & mat. & damped curvature proxy at $\theta_p$ \\
$C_f(\theta_0)\simeq\mathrm{diag}(H_f(\theta_0))$ & mat. & diagonal curvature proxy on $D_f$ \\
$\lambda>0$ & scalar & damping (Tikhonov) \\
\midrule
\multicolumn{3}{l}{\textbf{Masking, saliency, and indexing}}\\
$S_i(\cdot)$ & scalar & SNIP-style score for coordinate $i$ \\
$m^\star\in\{0,1\}^d$ & vec. & binary mask (1 = masked/zeroed)\\
$k=\|m^{\star}\|_0$ & int & mask budget (number of zeroed coords)\\
$M=\mathrm{supp}(m^{\star})$,\ $C=[d]\setminus M$ & set & masked index set; its complement \\
$E_M=[e_i]_{i\in M}$ & mat. & column selector (stacks basis vectors) \\
$\odot$ & op. & Hadamard (elementwise) product \\
\midrule
\multicolumn{3}{l}{\textbf{Updates and OBS compensation}}\\
$\delta w$ & vec. & total parameter update applied to $\theta_p$ \\
$\delta w_m=-\,\theta_p\odot m^{\star}$ & vec. & mask-induced removal (zeros $M$) \\
$\delta w_c$ & vec. & compensation on $C$ \\
$\displaystyle \delta w^{\star}$ & vec. & Group-OBS compensation ($w_{p,M}=E_M^\top\theta_p$) \\
$\lambda_M$ & vec. & KKT multipliers for Group-OBS constraints \\
$U(\theta_p;m^{\star})$ & op. & unlearning operator: $(\mathbf{1}-m^{\star})\odot\theta_p+\delta w^{\star}$ \\
\midrule
\multicolumn{3}{l}{\textbf{ZK verification (public vs.\ private artifacts)}}\\
$\Psi$ & pub. & traceability artifact (published mask $m^{\star}$ and $M$) \\
$\mathsf{Com}(\cdot)$ & pub. & binding commitment to an object (e.g., $\theta_p$, $C_p$, $\theta_u$) \\
% $\zeta$ & priv. & internal randomness (fixed/committed if present) \\
\bottomrule
\end{tabular}
\end{table*}

\end{document}